\documentstyle[12pt]{article}

\title{Monodisperse approximation in the metastable phase decay}

\author{V.Kurasov}

\date{Victor.Kurasov@pobox.spbu.ru}

\begin{document}

\maketitle

\begin{abstract}
A new simple method for the first order phase transition kinetics is suggested.
The  metastable phase consumption can be imagined in frames of the modisperse
approximation for the distribution of the droplets sizes. In all situations
of the metastable phase decay this approximation leads to negligible errors
in the total number of droplets appeared in the system. An evident advantage
of the presented method is the possibility to investigate the situation
of the metastable phase decay on several sorts of heterogeneous centers.
\end{abstract}

Attempts to give a theoretical description for the first order phase transition
appeared practically simultaneously with  creation of the classical nucleation
theory \cite{class}. An idea to give the global description  of the first
order phase transition was attractive and induced many publications (for
example, see Wakeshima \cite{Waka}, Segal' \cite{Segal}). But all of them
were intended only to estimate the main characteristics of the phase transition.
The time of the cut-off of the nucleation clearly observed in experiments
was adopted in these publications without any proper justification.
The first theoretical description with explicit determination of  time
behavior of  supersaturation was given in \cite{Kuni} where a homogeneous
condensation was investigated.
The method used in \cite{Kuni} was an iteration solution of  integral
equation of a substance balance.

To give a global picture of  phase transition one has to take into
account a presence of
active centers of  condensation.
The iteration method can be spread on  heterogeneous condensation
on similar centers (see \cite{Novoj}), but for  condensation
on  several types of centers one can not calculate iterations with a proper
accuracy (see also \cite{Novoj}). The system of condensation equations
becomes so complex that it can not be directly solved without simplifications.
As the result it would be rather attractive to suggest some simple approximations
which can be used in the complex situation of the condensation on several
types of centers. Certainly, this approximation has to be well based.

Here we shall suggest a monodisperse approximation of
the droplets size distribution to calculate the
number of molecules in the liquid phase. From the first point of view
this approximation is strange - we have already attracted attention to
the necessity to know the behavior of a supersaturation which determines
the form of the droplets size spectrum. But here we are going to show
that with an appropriate choice of the cut-off (which can be also justified)
one can give an adequate description of a nucleation period.

The monodisperse approximation presented here differs from the total monodisperse
approximation used in \cite{Kuni-het} for description of the intensive
consumption of the metastable phase and can not be applied during a nucleation
period. Here we use a special modification
of the mentioned approximation which allows to describe the period of
nucleation.

This publication is intended to give the main idea of the monodisperse
approximation which would be  clear for experimenters.
That's why we start from  situations of homogeneous
condensation and heterogeneous condensation on similar centers which can
be solved even without this approximation.
Some technical details are excluded to give the most clear version (for
example, a complete version of limit situations and monodisperse approximation
in  the intermediate situation is described in \cite{Multidecay}, the
overlapping of the intermediate and limit situations is analysed in \cite{Overlapping},
the transition of the special monodisperse approximation to the total
one is discussed in \cite{Ierarchy}).

We use the physical model of nucleation kinetics described in \cite{Novoj}.
It is rather  standard  but to avoid misunderstanding  we shall consider
\cite{Novoj}
 as the base of references.

\section{Homogeneous nucleation}

The condensation kinetics equation for the number  $G $
of the molecules in the liquid phase can be written in a
well known form \cite{Kuni}, \cite{Novoj}
\begin{equation}
\label{1}
G(z) = f \int_0^z dx (z-x)^3
\exp(-\Gamma G(x))
\end{equation}
where           parameter
$f$
is the amplitude          value of the
droplets sizes distribution
$\Gamma$
is some fixed positive parameter.

One can analyse behavior of subintegral function $g$ defined by
\begin{equation} \label{2}
   G(z)          = \int_0^z g(z,x) dx
\end{equation}
as a function of a size
 $\rho = z-x  $, $z$.
This function has the sense of the distribution of the number of molecules
in  droplets over their sizes $\rho$.

In some "moment" $z$ ( or $t(z) $)
it can be presented in the following  form
\begin{itemize}

\item
When $ \rho > z$
it is equal to zero (there are no droplets with such a big size)

\item
When $ \rho < 0 $
it
is also equal to zero (there aren't droplets with a negative size)

\item
At the intermediate $\rho $
it grows rather quickly with a growth of $ \rho$.
It is easy to note that it grows faster than $ \rho^3$ grows.

\end{itemize}

Really, if one takes into account that supersaturation decreases in time
then we get
$ g \sim \rho^3$.
But supersaturation falls in time and there aren't so many droplets of
the small size as of the big size.

As the result one can see that the function $ g $
as the function of $\rho$
has the sharp peak near  $\rho \approx z $.
This property takes place under the arbitrary $z $ (or $t(z) $).

The sharp peak of $ g$
allows to use for $g$ the monodisperse approximation - a representation
in the $\delta$-like form with a coordinate corresponding to  a position
of the peak of  function $g $, i.e.
$$
g \sim \delta(\rho = z)
$$
As the result one can state that the monodisperse approximation is based
now. But it is necessary to determine the number of droplets in this approximation.

It would be wrong to include the total number of already appeared droplets
in this peak. Really, in the spectrum of sizes there are many droplets
with small sizes. One can not describe these droplets as  containing the
same substance as  the droplets of a big size. It would be more correct
to exclude them from the substance balance. So, it is necessary to cut
off the small droplets. It can be done according to two recipes.

The first recipe is the differential one. One can note that during all
times which don't exceed the time of nucleation essentially the function
$g $
near maximum  is close to
$$
g_{appr} = f \rho^3
$$
This approximation corresponds to the constant value of  supersaturation.

One can cut off this approximation at a half of  amplitude value
(i.e. at a level $ f z^3 /
2$).
Then one can get for the width $ \Delta_{diff} z $
the following expression
$$
\Delta_{diff} z = (1-2^{-1/3}) z
$$

This cut off means that all droplets  $ \rho < z - \Delta_{diff}z   $
are excluded from consideration and all droplets
with $\rho >z - \Delta_{diff} z  $
are taken into account in a $\delta$-like peak.

The second recipe is the integral one. One can integrate $
g_{appr} $
and require that
$$
\int_0^z g_{appr} (z,x) dx = N  z^3
$$
An integration gives
$$
  \int_0^z g_{appr} (z,x) dx =  f \int_0^z (z-x)^3 dx = f \frac{z^4}{4}
$$
The width of spectrum is defined from condition that the number of droplets
has to be equal to the amplitude multiplied by the width of spectrum
$\Delta_{int} z $:
$$
N =          f \Delta_{int} z
$$
This gives the following expression
$$
\Delta_{int} z = z/4
$$

One can see that $\Delta_{diff}  z $  and $\Delta_{int}  z $
practically coincide. This shows the high selfconsistency of this approximation.
The second recipe will be more convenient for concrete calculations.

In fig.1 one can see the application of the monodisperse approximation
in the homogeneous case.

As the result one can say that all parameters of approximation are defined.
Now it will be used to solve (\ref{1}).

Instead of (\ref{1}) one can get
$$
G(z) = N(z/4) z^3
$$
where
$$
N(z/4) = f  z / 4
$$
is the number of droplets formed until  $t(z/4)   $.
This leads to
$$
G (z) = f z^4 / 4
$$
which coincides with the resulting iteration in the iteration method \cite{Kuni},
\cite{Novoj}. It is known (see \cite{Kuni}, \cite{Novoj}) that this expression
is very accurate which shows the effectiveness of the monodisperse approximation.
Here the cut off of the tail of the sizes spectrum compensates the unsymmetry
of the initial spectrum.

The main result of the nucleation process is the total number of the droplets
which can be found as
$$
  N_{tot} =          f
\int_0^{\infty} dx \exp(-\Gamma G(x)) dx
$$
or
$$
N_{tot} = f \int_0^{\infty} dx
\exp(-f \Gamma z^4 / 4) =
f^{3/4} \Gamma^{-1/4} D
$$
where
$$
D= \int_0^{\infty} \exp(-x^4/4) dx = 1.28
$$

The error of this expression is less than  two percents (it is the same as
in the iteration method).

\section{Heterogeneous condensation on similar centers}

The condensation equations system can be written in the following form
\cite{Novoj}
$$
G(z) = f \int_0^z dx (z-x)^3
\exp(-\Gamma G(x)) \theta (x)
$$
$$
\theta(z)
=
\exp( - b   \int_0^z \exp(-\Gamma G(x)) dx )
$$
with positive parameters $f$, $b$, $\Gamma$.
An appearance of a new function
$\theta$
which is a relative number of  free heterogeneous centers requires
the second equation.

The first equation of the system is rather analogous to the homogeneous
case.  The subintegral function here is also sharp. A function $\theta $
is a decreasing function of time according to the second equation of
the system.
Then the function $ g$
which is again determined by (\ref{2})
is more sharp than  in the homogeneous case.
As far as the supersaturation has to fall one can see that
$ g$
is more sharp than  $g_{appr}  $.
It allows to use here the monodisperse approximation for all $z $ or $t(z) $.

As the result the monodisperse approximation is based for heterogeneous
condensation. One needs here only the sharp peak of  $ g(\rho) $
which can be easily seen.

The successive application of the monodisperse approximation in the homogeneous
case shows that all droplets necessary for a
 metastable phase consumption at $ t(z)  $
were formed until  $ t(z/4) $.
In the heterogeneous case the exhaustion of heterogeneous centers increases
in time. So, all essential at  $ t(z)$
droplets were formed before
$ t(z/4)$.

At the same time the presence of a long tail in the situation of a weak
exhaustion of heterogeneous centers requires to cut off the spectrum
for the monodisperse approximation.
As far as the long tail is essential in the situation of a weak exhaustion
one has to cut off the spectrum by the same recipe as in the situation
of the homogeneous condensation: one has to exclude all droplets formed
after $z/4 $ which have the sizes  $\rho < z -
\Delta_{int} z = z - z/4 $.

One can see in fig. 2 the monodisperse approximation in the situation
of the heterogeneous condensation on similar centers.
The form of the spectrum in this situation is illustrated in fig. 3.
So, the way to construct approximation is known. Now one can turn to concrete
calculations.

The number of the droplets formed until $t(z/4) $
has to be calculated as
$$
N(z/4) = \frac{f}{b} (1 - \theta(z/4))
$$
An approximation for $G $
has the form
$$
G(z) = \frac{f}{b} (1- \theta(z/4)) z^3
$$

The total number of droplets can be determined as
$$
N_{tot} = \frac{f}{b} (1- \theta (\infty))
$$
or
$$
N_{tot} = \frac{f}{b}
(1-
\exp(
-b \int_0^{\infty} \exp(-\Gamma G(x)) dx
)
)
$$
or
$$
N_{tot} = \frac{f}{b}
(1-
\exp(
-b \int_0^{\infty} \exp(-\Gamma \frac{f}{b} (1-\theta(z/4)) z^3 ) dz
)
)
$$
or
$$
N_{tot} = \frac{f}{b}
(1-
\exp(
-b \int_0^{\infty} \exp(-\Gamma  \frac{f}{b}
(1-\exp
(-b
\int_0^{z/4}  \exp(-\Gamma
G(x) )  dx
)) z^3 ) dz
)
)
$$
The last expression has a rather complicate form. It contains several iterations
in a hidden form which ensures the high accuracy.

The last expression can be simplified. One of the possible recipes is
the following. One can note that an expression for $G  $
is necessary at $\Gamma G \sim 1 $.
Then $z  $
attains some values     $ \Delta_{\zeta} z  $.
But until $
\Delta_{\zeta} z / 4 $
 the value  $ \Gamma G $ is small and $
\exp(\Gamma G(z)) $
is close to unity. This leads to simplification of last expression which
can be written in the following form
$$
N_{tot} = \frac{f}{b}
(1-
\exp(
-b \int_0^{\infty} \exp(-\Gamma      \frac{f}{b}
(1-\exp
(-b z/4
)) z^3 ) dz
)
)
$$

Then one can fulfil calculation according to the last formula. The relative
error is less than two percents. Here it is a little bit greater than
in the homogeneous case because the form of initial spectrum  is changed and
there is no full compensation of the unsymmetry of spectrum and an exclusion
of the tail.
The relative error in the situation of heterogeneous condensation on similar
centers is drawn in fig. 4.

Now one can turn to explicit calculation of the integral in the last expression.
After the appropriate renormalization the subintegral function is more
sharp than
$ \exp(-x^3)$
and more smooth than $
\exp(-x^4) $.
Both these functions have a sharp back front of spectrum.
It allows to introduce the characteristic scale
$\Delta z$ by equation
$$
\frac{f}{b}
\Gamma
(1-\exp
(-b (\Delta z)/4
)) (\Delta z)^3 \approx 1
$$

Then
$$
\int_0^{\infty} \exp(-\Gamma \frac{f}{b}
(1-\exp
(-b z/4
)) z^3 ) dz =
\Delta z \frac{A+B}{2}
$$
where
$$
A = \int_0^{\infty} \exp(-x^3) dx = 0.89
$$
$$
B = \int_0^{\infty} \exp(-x^4) dx = 0.90
$$

Now the calculation is reduced to some algebraic manipulations. The error
of the last approximation is less than one percent.

As the result
$$
N_{tot} = \frac{f}{b}
(1-
\exp(
-b \Delta z \frac{A+B}{2}
)
)
$$

One can note that it is possible to formulate the recipe already in terms
of $
\Delta z $.
The long way is adopted here to give the most clear picture for the monodisperse
approximation.

\section{Nucleation on several types of heterogeneous centers}

The main advantage of  monodisperse approximation is the possibility
to use it for the condensation on the several types of centers.
The iteration procedure can not be applied in this case successfully. The
result of calculations according to \cite{Novoj} shows this fact explicitly.
The reason is the existence of the cross influence of the different types
of centers through vapor consumption.

In the condensation on similar heterogeneous centers in the situation
of exhaustion the influence of this phenomena on the vapor consumption
isn't important because in the situation of consumption the converging
force of the heterogeneous centers exhaustion is extremely high.
But in the situation with two types of heterogeneous centers the exhaustion
of the first type centers can have a certain influence on a vapor consumption
but the exhaustion of the second type centers is weak and there is no converging
force due to the weak exhaustion of the second type centers.

This effect in very thin and it can not be taken into account in the second
iteration. But one can not calculate the third iteration analytically
and this stops an application of  iterations. Really, this phenomena isn't
evident from the first point of view but it exits and leads to the error
of the second iteration in many times.

The application of the monodisperse approximation is based on the sharpness
of function
$  g $.
This property takes place already in this situation. So, there are no
objections to apply the monodisperse approximation here.

Here we shall reproduce the same formulas but with the lower indexes which
determine the sort of heterogeneous centers.

The system of condensation equations can be written in the following form
\cite{Multidecay}
$$
G_i(z) = f_i \int_0^z dx (z-x)^3
\exp(-\Gamma \sum_j G_j(x)) \theta_i (x)
$$
$$
\theta_i (z)
=
\exp( - b_i   \int_0^z \exp(-\Gamma \sum_j G_j(x)) dx )
$$
where the lower indexes
denote the sorts of centers.
This system can be seen by the direct generalization of
the one type case.

The subintegral function in the substance balance equations is also sharp.
As far as all  $\theta_i $
are the decreasing functions of arguments then the function
$ g$ defined by (\ref{2}) (with proper indexes)
is sharper than without the exhaustion of heterogeneous centers.
So, due to the supersaturationdecreasing
$ g$
is more sharp than  $g_{appr}  $.
It allows here to use the monodisperse approximation for all $z $ or $t(z) $.

As the result one can see that the monodisperse approximation
in this case is justified on the base of the sharpness of $ g(\rho) $.

The same properties as in the previous case can be also seen here. One
has to cut off the spectrum at $z/4$. Here all justifications are absolutely
same as in the previous section.
The characteristic situation for the nucleation on two types of heterogeneous
centers is drawn in fig.5.
As the result the way to construct the monodisperse approximation is known.
Now one can present calculations.

The number of the droplets formed until
$t(z/4) $
on the centers of sort  $i$
has to be calculated as
$$
N_i(z/4) = \frac{f_i}{b_i } (1 - \theta_i(z/4))
$$
An approximation for $G_i $
can be now presented as
$$
G_i (z) = \frac{f_i}{b_i } (1- \theta_i (z/4)) z^3
$$

The total number of droplets is defined as
$$
N_{i\ tot} = \frac{f_i}{b_i} (1- \theta_i (\infty))
$$
or
$$
N_{i\ tot} = \frac{f_i}{b_i}
(1-
\exp(
-b_i  \int_0^{\infty} \exp(-\Gamma \sum_j G_j (x)) dx
)
)
$$
or
$$
N_{i\ tot} = \frac{f_i}{b_i}
(1-
\exp(
-b_i \int_0^{\infty} \exp(-\Gamma \sum_j \frac{f_j}{b_j} (1-\theta_j(z/4)) z^3 ) dz
)
)
$$
or
\begin{eqnarray}
N_{i\ tot} = \frac{f_i}{b_i}
(1-
\exp(
-b_i \int_0^{\infty} \exp(-\Gamma
\nonumber
\\
\nonumber
\sum_j \frac{f_j}{b_j} (1-\exp
(-b_j
\int_0^{z/4}  \exp(-\Gamma
\sum_k G_k (x) )  dx
)) z^3 ) dz
)
)
\end{eqnarray}
Now one can simplify the last expression by the same way as in the one
type case.

Expressions for  $G_i  $
are essential at $\Gamma \sum_j G_j \sim 1 $.
Then $z  $ is near     $ \Delta_{\zeta} z  $.
Until $
\Delta_{\zeta} z / 4 $ the value  $ \Gamma \sum_j G_j $ is small  and
$
\exp(\Gamma \sum_j G_j (z)) $
is near unity.
It leads to
$$
N_{i\ tot} = \frac{f_i}{b_i}
 (1-
\exp(
-b_i \int_0^{\infty} \exp(-\Gamma
\sum_j \frac{f_j}{b_j} (1-\exp
(-b_j z/4
)) z^3 ) dz
)
)
$$

Now one can fulfil the calculations according the explicit formula.
The relative error of the last expression is less than five percents (here
it increases slightly due to the complex form of the spectrums on different
sorts.
The relative error in the number of droplets is drawn in fig. 6.
The calculation of the last integral is absolutely analogous to the previous
section. The subintegral function after renormalization lies between
$\exp(-x^3) $  and $
\exp(-x^4)$.
It allows to get the characteristic size
$\Delta z$ from
$$
\Gamma     \sum_j \frac{f_j}{b_j}
(1-\exp
(-b_j (\Delta z)/4
)) (\Delta z)^3 \approx 1
$$

Then
$$
\int_0^{\infty} \exp(-\Gamma     \sum_j \frac{f_j}{b_j}
(1-\exp
(-b_j z/4
)) z^3 ) dz =
\Delta z \frac{A+B}{2}
$$

The relative error of the last expression is less than one percent.

As the result
$$
N_{i\ tot} = \frac{f_i}{b_i}
(1-
\exp(
-b_i \Delta z \frac{A+B}{2}
)
)
$$
The formula is similar to the final expression in the previous section.
But parameters in the last formula have to be determined in another manner.

The physical sense of the last expression is the separate exhaustion of
heterogeneous centers. One sort of centers can influence on the other
sort only through a vapor consumption. This fact can be seen also in
the initial precise
system of the condensation equations.

\pagebreak

\begin{picture}(300,300)
\put(125,275){\line(0,-1){100}}
\put(125,275){.}
\put(125,275){\line(0,-1){100}}
\put(124,273){.}
\put(124,273){\line(0,-1){98}}
\put(123,270){.}
\put(123,270){\line(0,-1){95}}
\put(122,268){.}
\put(122,268){\line(0,-1){93}}
\put(121,265){.}
\put(121,265){\line(0,-1){90}}
\put(120,262){.}
\put(120,262){\line(0,-1){87}}
\put(119,259){.}
\put(119,259){\line(0,-1){84}}
\put(118,256){.}
\put(118,256){\line(0,-1){81}}
\put(117,254){.}
\put(117,254){\line(0,-1){79}}
\put(116,251){.}
\put(116,251){\line(0,-1){76}}
\put(115,249){.}
\put(115,249){\line(0,-1){74}}
\put(114,247){.}
\put(114,247){\line(0,-1){72}}
\put(113,244){.}
\put(113,244){\line(0,-1){69}}
\put(112,242){.}
\put(112,242){\line(0,-1){67}}
\put(111,240){.}
\put(111,240){\line(0,-1){65}}
\put(110,237){.}
\put(110,237){\line(0,-1){62}}
\put(109,235){.}
\put(109,235){\line(0,-1){60}}
\put(108,233){.}
\put(108,233){\line(0,-1){58}}
\put(107,231){.}
\put(107,231){\line(0,-1){56}}
\put(106,229){.}
\put(106,229){\line(0,-1){54}}
\put(105,227){.}
\put(105,227){\line(0,-1){52}}
\put(104,225){.}
\put(104,225){\line(0,-1){50}}
\put(103,223){.}
\put(103,223){\line(0,-1){48}}
\put(102,222){.}
\put(102,222){\line(0,-1){47}}
\put(101,220){.}
\put(101,220){\line(0,-1){45}}
\put(100,218){.}
\put(100,218){\line(0,-1){43}}
\put(99,216){\line(0,-1){41}}
\put(99,216){.}
\put(98,215){.}
\put(97,213){.}
\put(96,211){.}
\put(95,210){.}
\put(94,208){.}
\put(93,207){.}
\put(92,206){.}
\put(91,204){.}
\put(90,203){.}
\put(89,202){.}
\put(88,200){.}
\put(87,199){.}
\put(86,198){.}
\put(85,197){.}
\put(84,196){.}
\put(83,195){.}
\put(82,194){.}
\put(81,193){.}
\put(80,192){.}
\put(79,191){.}
\put(78,190){.}
\put(77,189){.}
\put(76,189){.}
\put(75,188){.}
\put(74,187){.}
\put(73,186){.}
\put(72,186){.}
\put(71,185){.}
\put(70,184){.}
\put(69,184){.}
\put(68,183){.}
\put(67,183){.}
\put(66,182){.}
\put(65,182){.}
\put(64,181){.}
\put(63,181){.}
\put(62,180){.}
\put(61,180){.}
\put(60,179){.}
\put(59,179){.}
\put(58,179){.}
\put(57,178){.}
\put(56,178){.}
\put(55,178){.}
\put(54,178){.}
\put(53,177){.}
\put(52,177){.}
\put(51,177){.}
\put(50,177){.}
\put(49,176){.}
\put(48,176){.}
\put(47,176){.}
\put(46,176){.}
\put(45,176){.}
\put(44,176){.}
\put(43,176){.}
\put(42,176){.}
\put(41,175){.}
\put(40,175){.}
\put(39,175){.}
\put(38,175){.}
\put(37,175){.}
\put(36,175){.}
\put(35,175){.}
\put(34,175){.}
\put(33,175){.}
\put(32,175){.}
\put(31,175){.}
\put(30,175){.}
\put(29,175){.}
\put(28,175){.}
\put(27,175){.}
\put(26,175){.}
\put(25,175){.}
\put(25,175){\vector(0,1){110}}
\put(25,175){\vector(1,0){110}}
\put(125,173){\line(0,1){4}}
\put(22,275){\line(1,0){4}}
\put(125,168){$z$}
\put(3,275){$f z^3$}
\put(135,168){$\rho$}
\put(18,290){$g$}
\put(75,163){A}
\put(275,275){\line(0,-1){100}}
\put(275,275){.}
\put(275,275){\line(0,-1){100}}
\put(274,273){.}
\put(274,273){\line(0,-1){98}}
\put(273,270){.}
\put(273,270){\line(0,-1){95}}
\put(272,268){.}
\put(272,268){\line(0,-1){93}}
\put(271,265){.}
\put(271,265){\line(0,-1){90}}
\put(270,262){.}
\put(270,262){\line(0,-1){87}}
\put(269,259){.}
\put(269,259){\line(0,-1){84}}
\put(268,257){.}
\put(268,257){\line(0,-1){82}}
\put(267,254){.}
\put(267,254){\line(0,-1){79}}
\put(266,252){.}
\put(266,252){\line(0,-1){77}}
\put(265,249){.}
\put(265,249){\line(0,-1){74}}
\put(264,247){.}
\put(264,247){\line(0,-1){72}}
\put(263,244){.}
\put(263,244){\line(0,-1){69}}
\put(262,242){.}
\put(262,242){\line(0,-1){67}}
\put(261,240){.}
\put(261,240){\line(0,-1){65}}
\put(260,237){.}
\put(260,237){\line(0,-1){62}}
\put(259,235){.}
\put(259,235){\line(0,-1){60}}
\put(258,233){.}
\put(258,233){\line(0,-1){58}}
\put(257,231){.}
\put(257,231){\line(0,-1){56}}
\put(256,229){.}
\put(256,229){\line(0,-1){54}}
\put(255,227){.}
\put(255,227){\line(0,-1){52}}
\put(254,225){.}
\put(254,225){\line(0,-1){50}}
\put(253,223){.}
\put(253,223){\line(0,-1){48}}
\put(252,221){.}
\put(252,221){\line(0,-1){46}}
\put(251,220){.}
\put(251,220){\line(0,-1){45}}
\put(250,218){.}
\put(250,218){\line(0,-1){43}}
\put(249,216){\line(0,-1){41}}
\put(249,216){.}
\put(248,214){.}
\put(247,213){.}
\put(246,211){.}
\put(245,210){.}
\put(244,208){.}
\put(243,207){.}
\put(242,205){.}
\put(241,204){.}
\put(240,203){.}
\put(239,201){.}
\put(238,200){.}
\put(237,199){.}
\put(236,198){.}
\put(235,197){.}
\put(234,195){.}
\put(233,194){.}
\put(232,193){.}
\put(231,192){.}
\put(230,191){.}
\put(229,190){.}
\put(228,190){.}
\put(227,189){.}
\put(226,188){.}
\put(225,187){.}
\put(224,186){.}
\put(223,186){.}
\put(222,185){.}
\put(221,184){.}
\put(220,184){.}
\put(219,183){.}
\put(218,182){.}
\put(217,182){.}
\put(216,181){.}
\put(215,181){.}
\put(214,180){.}
\put(213,180){.}
\put(212,179){.}
\put(211,179){.}
\put(210,179){.}
\put(209,178){.}
\put(208,178){.}
\put(207,178){.}
\put(206,177){.}
\put(205,177){.}
\put(204,177){.}
\put(203,177){.}
\put(202,177){.}
\put(201,176){.}
\put(200,176){.}
\put(199,176){.}
\put(198,176){.}
\put(197,176){.}
\put(196,176){.}
\put(195,176){.}
\put(194,175){.}
\put(193,175){.}
\put(192,175){.}
\put(191,175){.}
\put(190,175){.}
\put(189,175){.}
\put(188,175){.}
\put(187,175){.}
\put(186,175){.}
\put(185,175){.}
\put(184,175){.}
\put(183,175){.}
\put(182,175){.}
\put(181,175){.}
\put(180,175){.}
\put(179,175){.}
\put(178,175){.}
\put(177,175){.}
\put(176,175){.}
\put(175,175){.}
\put(175,175){\vector(0,1){110}}
\put(175,175){\vector(1,0){110}}
\put(275,173){\line(0,1){4}}
\put(172,275){\line(1,0){4}}
\put(225,163){B}
\put(275,168){$z$}
\put(153,275){$f z^3$}
\put(285,168){$\rho$}
\put(168,290){$g$}
\put(125,125){\line(0,-1){100}}
\put(125,125){.}
\put(125,125){\line(0,-1){100}}
\put(124,123){.}
\put(124,123){\line(0,-1){98}}
\put(123,121){.}
\put(123,121){\line(0,-1){96}}
\put(122,118){.}
\put(122,118){\line(0,-1){93}}
\put(121,115){.}
\put(121,115){\line(0,-1){90}}
\put(120,112){.}
\put(120,112){\line(0,-1){87}}
\put(119,109){.}
\put(119,109){\line(0,-1){84}}
\put(118,107){.}
\put(118,107){\line(0,-1){82}}
\put(117,104){.}
\put(117,104){\line(0,-1){79}}
\put(116,102){.}
\put(116,102){\line(0,-1){77}}
\put(115,99){.}
\put(115,99){\line(0,-1){74}}
\put(114,97){.}
\put(114,97){\line(0,-1){72}}
\put(113,94){.}
\put(113,94){\line(0,-1){69}}
\put(112,92){.}
\put(112,92){\line(0,-1){67}}
\put(111,89){.}
\put(111,89){\line(0,-1){64}}
\put(110,87){.}
\put(110,87){\line(0,-1){62}}
\put(109,85){.}
\put(109,85){\line(0,-1){60}}
\put(108,83){.}
\put(108,83){\line(0,-1){58}}
\put(107,80){.}
\put(107,80){\line(0,-1){55}}
\put(106,78){.}
\put(106,78){\line(0,-1){53}}
\put(105,76){.}
\put(105,76){\line(0,-1){51}}
\put(104,74){.}
\put(104,74){\line(0,-1){49}}
\put(103,72){.}
\put(103,72){\line(0,-1){47}}
\put(102,70){.}
\put(102,70){\line(0,-1){45}}
\put(101,68){.}
\put(101,68){\line(0,-1){43}}
\put(100,66){.}
\put(100,66){\line(0,-1){41}}
\put(99,64){\line(0,-1){39}}
\put(99,64){.}
\put(98,62){.}
\put(97,60){.}
\put(96,58){.}
\put(95,56){.}
\put(94,54){.}
\put(93,52){.}
\put(92,51){.}
\put(91,49){.}
\put(90,47){.}
\put(89,46){.}
\put(88,44){.}
\put(87,43){.}
\put(86,41){.}
\put(85,40){.}
\put(84,39){.}
\put(83,37){.}
\put(82,36){.}
\put(81,35){.}
\put(80,34){.}
\put(79,33){.}
\put(78,32){.}
\put(77,31){.}
\put(76,31){.}
\put(75,30){.}
\put(74,29){.}
\put(73,29){.}
\put(72,28){.}
\put(71,28){.}
\put(70,27){.}
\put(69,27){.}
\put(68,27){.}
\put(67,26){.}
\put(66,26){.}
\put(65,26){.}
\put(64,26){.}
\put(63,26){.}
\put(62,25){.}
\put(61,25){.}
\put(60,25){.}
\put(59,25){.}
\put(58,25){.}
\put(57,25){.}
\put(56,25){.}
\put(55,25){.}
\put(54,25){.}
\put(53,25){.}
\put(52,25){.}
\put(51,25){.}
\put(50,25){.}
\put(49,25){.}
\put(48,25){.}
\put(47,25){.}
\put(46,25){.}
\put(45,25){.}
\put(44,25){.}
\put(43,25){.}
\put(42,25){.}
\put(41,25){.}
\put(40,25){.}
\put(39,25){.}
\put(38,25){.}
\put(37,25){.}
\put(36,25){.}
\put(35,25){.}
\put(34,25){.}
\put(33,25){.}
\put(32,25){.}
\put(31,25){.}
\put(30,25){.}
\put(29,25){.}
\put(28,25){.}
\put(27,25){.}
\put(26,25){.}
\put(25,25){.}
\put(25,25){\vector(0,1){110}}
\put(25,25){\vector(1,0){110}}
\put(125,23){\line(0,1){4}}
\put(22,125){\line(1,0){4}}
\put(125,18){$z$}
\put(3,125){$f z^3$}
\put(135,18){$\rho$}
\put(18,140){$g$}
\put(75,13){C}
\put(275,125){\line(0,-1){100}}
\put(275,125){.}
\put(275,125){\line(0,-1){100}}
\put(274,123){.}
\put(274,123){\line(0,-1){98}}
\put(273,121){.}
\put(273,121){\line(0,-1){96}}
\put(272,118){.}
\put(272,118){\line(0,-1){93}}
\put(271,115){.}
\put(271,115){\line(0,-1){90}}
\put(270,112){.}
\put(270,112){\line(0,-1){87}}
\put(269,109){.}
\put(269,109){\line(0,-1){84}}
\put(268,106){.}
\put(268,106){\line(0,-1){81}}
\put(267,103){.}
\put(267,103){\line(0,-1){78}}
\put(266,101){.}
\put(266,101){\line(0,-1){76}}
\put(265,98){.}
\put(265,98){\line(0,-1){73}}
\put(264,94){.}
\put(264,94){\line(0,-1){69}}
\put(263,91){.}
\put(263,91){\line(0,-1){66}}
\put(262,88){.}
\put(262,88){\line(0,-1){63}}
\put(261,84){.}
\put(261,84){\line(0,-1){59}}
\put(260,81){.}
\put(260,81){\line(0,-1){56}}
\put(259,77){.}
\put(259,77){\line(0,-1){52}}
\put(258,73){.}
\put(258,73){\line(0,-1){48}}
\put(257,69){.}
\put(257,69){\line(0,-1){44}}
\put(256,65){.}
\put(256,65){\line(0,-1){40}}
\put(255,61){.}
\put(255,61){\line(0,-1){36}}
\put(254,57){.}
\put(254,57){\line(0,-1){32}}
\put(253,53){.}
\put(253,53){\line(0,-1){28}}
\put(252,49){.}
\put(252,49){\line(0,-1){24}}
\put(251,45){.}
\put(251,45){\line(0,-1){20}}
\put(250,42){.}
\put(250,42){\line(0,-1){17}}
\put(249,39){\line(0,-1){14}}
\put(249,39){.}
\put(248,36){.}
\put(247,34){.}
\put(246,32){.}
\put(245,30){.}
\put(244,29){.}
\put(243,28){.}
\put(242,27){.}
\put(241,26){.}
\put(240,26){.}
\put(239,26){.}
\put(238,25){.}
\put(237,25){.}
\put(236,25){.}
\put(235,25){.}
\put(234,25){.}
\put(233,25){.}
\put(232,25){.}
\put(231,25){.}
\put(230,25){.}
\put(229,25){.}
\put(228,25){.}
\put(227,25){.}
\put(226,25){.}
\put(225,25){.}
\put(224,25){.}
\put(223,25){.}
\put(222,25){.}
\put(221,25){.}
\put(220,25){.}
\put(219,25){.}
\put(175,25){\vector(0,1){110}}
\put(175,25){\vector(1,0){110}}
\put(275,23){\line(0,1){4}}
\put(172,125){\line(1,0){4}}
\put(275,18){$z$}
\put(153,125){$f z^3$}
\put(285,18){$\rho$}
\put(168,140){$g$}
\put(225,13){D}
\put(0,0){\line(1,0){300}}
\put(0,0){\line(0,1){300}}
\put(0,300){\line(1,0){300}}
\put(300,0){\line(0,1){300}}
\end{picture}

\begin{center}

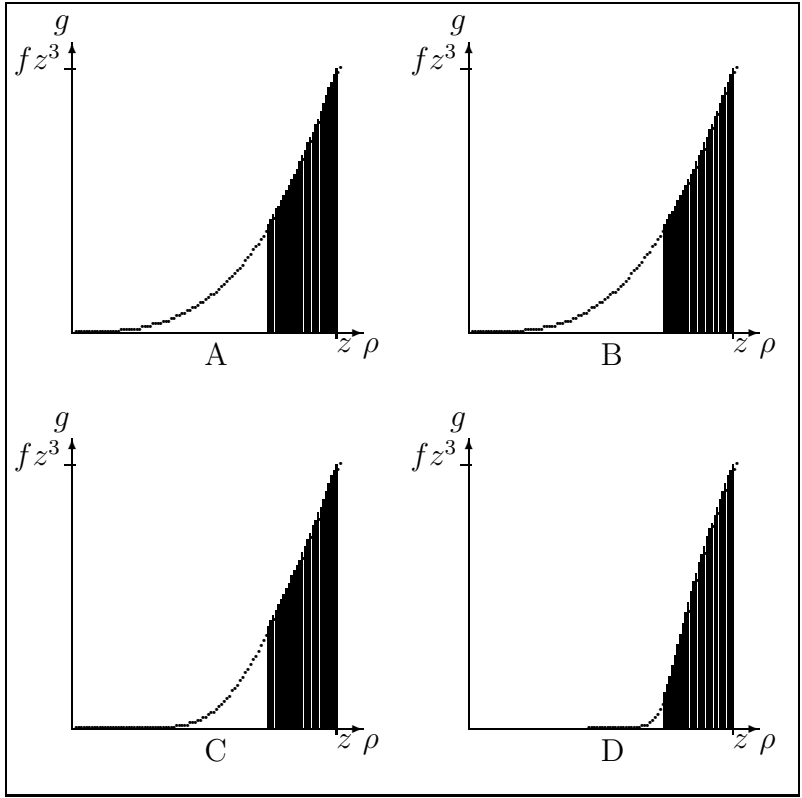
{\it Fig.1}

\end{center}

{\it   Monodisperse approximation in homogeneous condensation. Here one
can see four pictures for different periods of time (or for different
values of $z$. One can introduce $\Delta z$ according to $\Gamma G(\Delta
z) = 1$ and it will be the characteristic scale of the supersaturation
fall. In part "A" $z=\Delta z/2$, in part "B" $z=\Delta z$, in part "C"
$z= 3 \Delta z / 2$, in part "D" $z= 2 \Delta z$. One can see that the
spectrums in part "A" and part "B" are practically the same. It corresponds
to the property of the similarity of spectrums until the end of the nucleation
period.  }

\pagebreak

\begin{picture}(300,300)
\put(125,275){\line(0,-1){100}}
\put(125,275){.}
\put(125,275){\line(0,-1){100}}
\put(124,273){.}
\put(124,273){\line(0,-1){98}}
\put(123,269){.}
\put(123,269){\line(0,-1){94}}
\put(122,266){.}
\put(122,266){\line(0,-1){91}}
\put(121,263){.}
\put(121,263){\line(0,-1){88}}
\put(120,259){.}
\put(120,259){\line(0,-1){84}}
\put(119,256){.}
\put(119,256){\line(0,-1){81}}
\put(118,253){.}
\put(118,253){\line(0,-1){78}}
\put(117,250){.}
\put(117,250){\line(0,-1){75}}
\put(116,247){.}
\put(116,247){\line(0,-1){72}}
\put(115,244){.}
\put(115,244){\line(0,-1){69}}
\put(114,241){.}
\put(114,241){\line(0,-1){66}}
\put(113,239){.}
\put(113,239){\line(0,-1){64}}
\put(112,236){.}
\put(112,236){\line(0,-1){61}}
\put(111,234){.}
\put(111,234){\line(0,-1){59}}
\put(110,231){.}
\put(110,231){\line(0,-1){56}}
\put(109,229){.}
\put(109,229){\line(0,-1){54}}
\put(108,227){.}
\put(108,227){\line(0,-1){52}}
\put(107,224){.}
\put(107,224){\line(0,-1){49}}
\put(106,222){.}
\put(106,222){\line(0,-1){47}}
\put(105,220){.}
\put(105,220){\line(0,-1){45}}
\put(104,218){.}
\put(104,218){\line(0,-1){43}}
\put(103,216){.}
\put(103,216){\line(0,-1){41}}
\put(102,215){.}
\put(102,215){\line(0,-1){40}}
\put(101,213){.}
\put(101,213){\line(0,-1){38}}
\put(100,211){.}
\put(100,211){\line(0,-1){36}}
\put(99,209){\line(0,-1){34}}
\put(99,209){.}
\put(98,208){.}
\put(97,206){.}
\put(96,205){.}
\put(95,203){.}
\put(94,202){.}
\put(93,201){.}
\put(92,199){.}
\put(91,198){.}
\put(90,197){.}
\put(89,196){.}
\put(88,195){.}
\put(87,194){.}
\put(86,193){.}
\put(85,192){.}
\put(84,191){.}
\put(83,190){.}
\put(82,189){.}
\put(81,188){.}
\put(80,187){.}
\put(79,187){.}
\put(78,186){.}
\put(77,185){.}
\put(76,185){.}
\put(75,184){.}
\put(74,183){.}
\put(73,183){.}
\put(72,182){.}
\put(71,182){.}
\put(70,181){.}
\put(69,181){.}
\put(68,180){.}
\put(67,180){.}
\put(66,180){.}
\put(65,179){.}
\put(64,179){.}
\put(63,179){.}
\put(62,178){.}
\put(61,178){.}
\put(60,178){.}
\put(59,178){.}
\put(58,177){.}
\put(57,177){.}
\put(56,177){.}
\put(55,177){.}
\put(54,177){.}
\put(53,176){.}
\put(52,176){.}
\put(51,176){.}
\put(50,176){.}
\put(49,176){.}
\put(48,176){.}
\put(47,176){.}
\put(46,176){.}
\put(45,175){.}
\put(44,175){.}
\put(43,175){.}
\put(42,175){.}
\put(41,175){.}
\put(40,175){.}
\put(39,175){.}
\put(38,175){.}
\put(37,175){.}
\put(36,175){.}
\put(35,175){.}
\put(34,175){.}
\put(33,175){.}
\put(32,175){.}
\put(31,175){.}
\put(30,175){.}
\put(29,175){.}
\put(28,175){.}
\put(27,175){.}
\put(26,175){.}
\put(25,175){.}
\put(25,175){\vector(0,1){110}}
\put(25,175){\vector(1,0){110}}
\put(125,173){\line(0,1){4}}
\put(22,275){\line(1,0){4}}
\put(125,168){$z$}
\put(3,275){$f z^3$}
\put(135,168){$\rho$}
\put(18,290){$g$}
\put(75,163){A}
\put(275,275){\line(0,-1){100}}
\put(275,275){.}
\put(275,275){\line(0,-1){100}}
\put(274,273){.}
\put(274,273){\line(0,-1){98}}
\put(273,268){.}
\put(273,268){\line(0,-1){93}}
\put(272,264){.}
\put(272,264){\line(0,-1){89}}
\put(271,261){.}
\put(271,261){\line(0,-1){86}}
\put(270,257){.}
\put(270,257){\line(0,-1){82}}
\put(269,253){.}
\put(269,253){\line(0,-1){78}}
\put(268,249){.}
\put(268,249){\line(0,-1){74}}
\put(267,246){.}
\put(267,246){\line(0,-1){71}}
\put(266,243){.}
\put(266,243){\line(0,-1){68}}
\put(265,240){.}
\put(265,240){\line(0,-1){65}}
\put(264,237){.}
\put(264,237){\line(0,-1){62}}
\put(263,234){.}
\put(263,234){\line(0,-1){59}}
\put(262,231){.}
\put(262,231){\line(0,-1){56}}
\put(261,228){.}
\put(261,228){\line(0,-1){53}}
\put(260,226){.}
\put(260,226){\line(0,-1){51}}
\put(259,223){.}
\put(259,223){\line(0,-1){48}}
\put(258,221){.}
\put(258,221){\line(0,-1){46}}
\put(257,219){.}
\put(257,219){\line(0,-1){44}}
\put(256,217){.}
\put(256,217){\line(0,-1){42}}
\put(255,214){.}
\put(255,214){\line(0,-1){39}}
\put(254,212){.}
\put(254,212){\line(0,-1){37}}
\put(253,211){.}
\put(253,211){\line(0,-1){36}}
\put(252,209){.}
\put(252,209){\line(0,-1){34}}
\put(251,207){.}
\put(251,207){\line(0,-1){32}}
\put(250,205){.}
\put(250,205){\line(0,-1){30}}
\put(249,204){\line(0,-1){29}}
\put(249,204){.}
\put(248,202){.}
\put(247,201){.}
\put(246,199){.}
\put(245,198){.}
\put(244,197){.}
\put(243,195){.}
\put(242,194){.}
\put(241,193){.}
\put(240,192){.}
\put(239,191){.}
\put(238,190){.}
\put(237,189){.}
\put(236,188){.}
\put(235,187){.}
\put(234,187){.}
\put(233,186){.}
\put(232,185){.}
\put(231,184){.}
\put(230,184){.}
\put(229,183){.}
\put(228,183){.}
\put(227,182){.}
\put(226,182){.}
\put(225,181){.}
\put(224,181){.}
\put(223,180){.}
\put(222,180){.}
\put(221,179){.}
\put(220,179){.}
\put(219,179){.}
\put(218,178){.}
\put(217,178){.}
\put(216,178){.}
\put(215,178){.}
\put(214,177){.}
\put(213,177){.}
\put(212,177){.}
\put(211,177){.}
\put(210,177){.}
\put(209,176){.}
\put(208,176){.}
\put(207,176){.}
\put(206,176){.}
\put(205,176){.}
\put(204,176){.}
\put(203,176){.}
\put(202,176){.}
\put(201,176){.}
\put(200,175){.}
\put(199,175){.}
\put(198,175){.}
\put(197,175){.}
\put(196,175){.}
\put(195,175){.}
\put(194,175){.}
\put(193,175){.}
\put(192,175){.}
\put(191,175){.}
\put(190,175){.}
\put(189,175){.}
\put(188,175){.}
\put(187,175){.}
\put(186,175){.}
\put(185,175){.}
\put(184,175){.}
\put(183,175){.}
\put(182,175){.}
\put(181,175){.}
\put(180,175){.}
\put(179,175){.}
\put(178,175){.}
\put(177,175){.}
\put(176,175){.}
\put(175,175){.}
\put(175,175){\vector(0,1){110}}
\put(175,175){\vector(1,0){110}}
\put(275,173){\line(0,1){4}}
\put(172,275){\line(1,0){4}}
\put(225,163){B}
\put(275,168){$z$}
\put(153,275){$f z^3$}
\put(285,168){$\rho$}
\put(168,290){$g$}
\put(125,125){\line(0,-1){100}}
\put(125,125){.}
\put(125,125){\line(0,-1){100}}
\put(124,122){.}
\put(124,122){\line(0,-1){97}}
\put(123,117){.}
\put(123,117){\line(0,-1){92}}
\put(122,111){.}
\put(122,111){\line(0,-1){86}}
\put(121,106){.}
\put(121,106){\line(0,-1){81}}
\put(120,102){.}
\put(120,102){\line(0,-1){77}}
\put(119,97){.}
\put(119,97){\line(0,-1){72}}
\put(118,93){.}
\put(118,93){\line(0,-1){68}}
\put(117,89){.}
\put(117,89){\line(0,-1){64}}
\put(116,85){.}
\put(116,85){\line(0,-1){60}}
\put(115,82){.}
\put(115,82){\line(0,-1){57}}
\put(114,78){.}
\put(114,78){\line(0,-1){53}}
\put(113,75){.}
\put(113,75){\line(0,-1){50}}
\put(112,72){.}
\put(112,72){\line(0,-1){47}}
\put(111,69){.}
\put(111,69){\line(0,-1){44}}
\put(110,66){.}
\put(110,66){\line(0,-1){41}}
\put(109,64){.}
\put(109,64){\line(0,-1){39}}
\put(108,61){.}
\put(108,61){\line(0,-1){36}}
\put(107,59){.}
\put(107,59){\line(0,-1){34}}
\put(106,57){.}
\put(106,57){\line(0,-1){32}}
\put(105,54){.}
\put(105,54){\line(0,-1){29}}
\put(104,52){.}
\put(104,52){\line(0,-1){27}}
\put(103,51){.}
\put(103,51){\line(0,-1){26}}
\put(102,49){.}
\put(102,49){\line(0,-1){24}}
\put(101,47){.}
\put(101,47){\line(0,-1){22}}
\put(100,46){.}
\put(100,46){\line(0,-1){21}}
\put(99,44){\line(0,-1){19}}
\put(99,44){.}
\put(98,43){.}
\put(97,41){.}
\put(96,40){.}
\put(95,39){.}
\put(94,38){.}
\put(93,37){.}
\put(92,36){.}
\put(91,35){.}
\put(90,34){.}
\put(89,33){.}
\put(88,32){.}
\put(87,32){.}
\put(86,31){.}
\put(85,31){.}
\put(84,30){.}
\put(83,30){.}
\put(82,29){.}
\put(81,29){.}
\put(80,28){.}
\put(79,28){.}
\put(78,28){.}
\put(77,27){.}
\put(76,27){.}
\put(75,27){.}
\put(74,27){.}
\put(73,26){.}
\put(72,26){.}
\put(71,26){.}
\put(70,26){.}
\put(69,26){.}
\put(68,26){.}
\put(67,26){.}
\put(66,25){.}
\put(65,25){.}
\put(64,25){.}
\put(63,25){.}
\put(62,25){.}
\put(61,25){.}
\put(60,25){.}
\put(59,25){.}
\put(58,25){.}
\put(57,25){.}
\put(56,25){.}
\put(55,25){.}
\put(54,25){.}
\put(53,25){.}
\put(52,25){.}
\put(51,25){.}
\put(50,25){.}
\put(49,25){.}
\put(48,25){.}
\put(47,25){.}
\put(46,25){.}
\put(45,25){.}
\put(44,25){.}
\put(43,25){.}
\put(42,25){.}
\put(41,25){.}
\put(40,25){.}
\put(39,25){.}
\put(38,25){.}
\put(37,25){.}
\put(36,25){.}
\put(35,25){.}
\put(34,25){.}
\put(33,25){.}
\put(32,25){.}
\put(31,25){.}
\put(30,25){.}
\put(29,25){.}
\put(28,25){.}
\put(27,25){.}
\put(26,25){.}
\put(25,25){.}
\put(25,25){\vector(0,1){110}}
\put(25,25){\vector(1,0){110}}
\put(125,23){\line(0,1){4}}
\put(22,125){\line(1,0){4}}
\put(125,18){$z$}
\put(3,125){$f z^3$}
\put(135,18){$\rho$}
\put(18,140){$g$}
\put(75,13){C}
\put(275,125){\line(0,-1){100}}
\put(275,125){.}
\put(275,125){\line(0,-1){100}}
\put(274,121){.}
\put(274,121){\line(0,-1){96}}
\put(273,113){.}
\put(273,113){\line(0,-1){88}}
\put(272,105){.}
\put(272,105){\line(0,-1){80}}
\put(271,99){.}
\put(271,99){\line(0,-1){74}}
\put(270,92){.}
\put(270,92){\line(0,-1){67}}
\put(269,87){.}
\put(269,87){\line(0,-1){62}}
\put(268,81){.}
\put(268,81){\line(0,-1){56}}
\put(267,76){.}
\put(267,76){\line(0,-1){51}}
\put(266,72){.}
\put(266,72){\line(0,-1){47}}
\put(265,68){.}
\put(265,68){\line(0,-1){43}}
\put(264,64){.}
\put(264,64){\line(0,-1){39}}
\put(263,60){.}
\put(263,60){\line(0,-1){35}}
\put(262,56){.}
\put(262,56){\line(0,-1){31}}
\put(261,53){.}
\put(261,53){\line(0,-1){28}}
\put(260,50){.}
\put(260,50){\line(0,-1){25}}
\put(259,47){.}
\put(259,47){\line(0,-1){22}}
\put(258,45){.}
\put(258,45){\line(0,-1){20}}
\put(257,43){.}
\put(257,43){\line(0,-1){18}}
\put(256,41){.}
\put(256,41){\line(0,-1){16}}
\put(255,39){.}
\put(255,39){\line(0,-1){14}}
\put(254,37){.}
\put(254,37){\line(0,-1){12}}
\put(253,35){.}
\put(253,35){\line(0,-1){10}}
\put(252,34){.}
\put(252,34){\line(0,-1){9}}
\put(251,32){.}
\put(251,32){\line(0,-1){7}}
\put(250,31){.}
\put(250,31){\line(0,-1){6}}
\put(249,30){\line(0,-1){5}}
\put(249,30){.}
\put(248,29){.}
\put(247,28){.}
\put(246,28){.}
\put(245,27){.}
\put(244,27){.}
\put(243,26){.}
\put(242,26){.}
\put(241,26){.}
\put(240,26){.}
\put(239,25){.}
\put(238,25){.}
\put(237,25){.}
\put(236,25){.}
\put(235,25){.}
\put(234,25){.}
\put(233,25){.}
\put(232,25){.}
\put(231,25){.}
\put(230,25){.}
\put(229,25){.}
\put(228,25){.}
\put(227,25){.}
\put(226,25){.}
\put(225,25){.}
\put(224,25){.}
\put(223,25){.}
\put(222,25){.}
\put(221,25){.}
\put(220,25){.}
\put(219,25){.}
\put(218,25){.}
\put(217,25){.}
\put(216,25){.}
\put(215,25){.}
\put(214,25){.}
\put(213,25){.}
\put(175,25){\vector(0,1){110}}
\put(175,25){\vector(1,0){110}}
\put(275,23){\line(0,1){4}}
\put(172,125){\line(1,0){4}}
\put(275,18){$z$}
\put(153,125){$f z^3$}
\put(285,18){$\rho$}
\put(168,140){$g$}
\put(225,13){D}
\put(0,0){\line(1,0){300}}
\put(0,0){\line(0,1){300}}
\put(0,300){\line(1,0){300}}
\put(300,0){\line(0,1){300}}
\end{picture}

\begin{center}

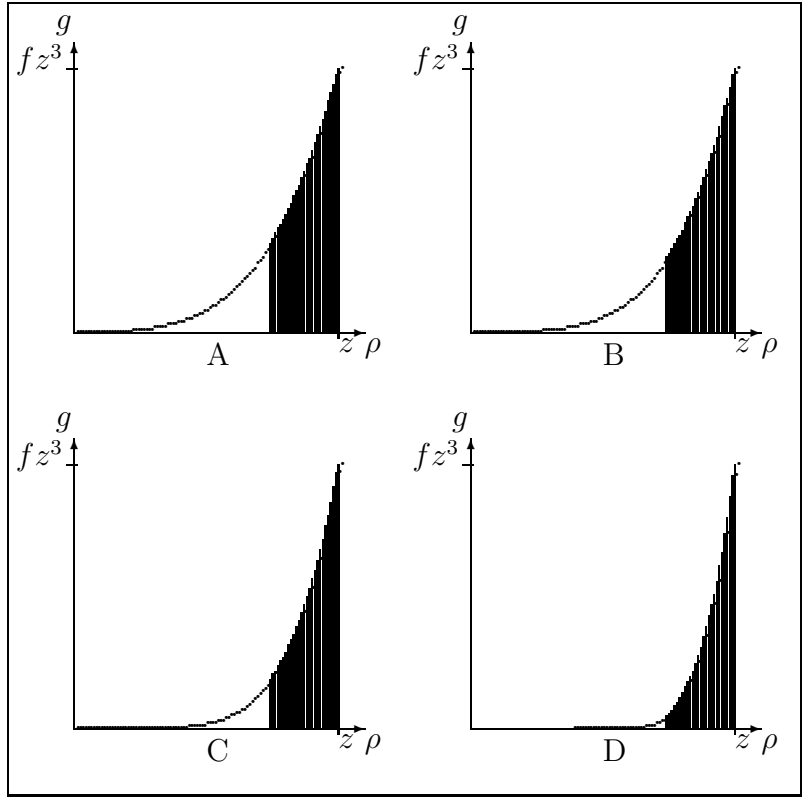
{\it Fig.2}

\end{center}

{\it   Monodisperse approximation in condensation on the similar centers.
The value $\Delta z$ is the same as in the previous figure (i.e. it is
determined without the exhaustion of centers). Now  $\Gamma G(\Delta
z) < 1$ and $\Delta z$  will be the characteristic scale of the supersaturation
fall in the situation without exhaustion.
In part "A" $z=\Delta z/2$, in part "B" $z=\Delta z$, in part "C"
$z= 3 \Delta z / 2$, in part "D" $z= 2 \Delta z$. One can see that the
spectrums in part "A" and part "B" aren't similar.
Now all spectrums are more sharp than in the homogeneous case. }

\pagebreak

\begin{picture}(400,300)
\put(325,225){\line(0,-1){200}}
\put(325,225){.}
\put(324,224){.}
\put(323,223){.}
\put(322,222){.}
\put(321,220){.}
\put(320,219){.}
\put(319,217){.}
\put(318,216){.}
\put(317,215){.}
\put(316,213){.}
\put(315,212){.}
\put(314,211){.}
\put(313,209){.}
\put(312,208){.}
\put(311,207){.}
\put(310,205){.}
\put(309,204){.}
\put(308,203){.}
\put(307,202){.}
\put(306,200){.}
\put(305,199){.}
\put(304,198){.}
\put(303,197){.}
\put(302,196){.}
\put(301,194){.}
\put(300,193){.}
\put(299,192){.}
\put(298,191){.}
\put(297,190){.}
\put(296,188){.}
\put(295,187){.}
\put(294,186){.}
\put(293,185){.}
\put(292,184){.}
\put(291,183){.}
\put(290,181){.}
\put(289,180){.}
\put(288,179){.}
\put(287,178){.}
\put(286,177){.}
\put(285,176){.}
\put(284,175){.}
\put(283,174){.}
\put(282,173){.}
\put(281,172){.}
\put(280,170){.}
\put(279,169){.}
\put(278,168){.}
\put(277,167){.}
\put(276,166){.}
\put(275,165){.}
\put(274,164){.}
\put(273,163){.}
\put(272,162){.}
\put(271,161){.}
\put(270,160){.}
\put(269,159){.}
\put(268,158){.}
\put(267,157){.}
\put(266,156){.}
\put(265,155){.}
\put(264,154){.}
\put(263,153){.}
\put(262,152){.}
\put(261,151){.}
\put(260,150){.}
\put(259,149){.}
\put(258,148){.}
\put(257,148){.}
\put(256,147){.}
\put(255,146){.}
\put(254,145){.}
\put(253,144){.}
\put(252,143){.}
\put(251,142){.}
\put(250,141){.}
\put(249,140){.}
\put(248,139){.}
\put(247,138){.}
\put(246,137){.}
\put(245,137){.}
\put(244,136){.}
\put(243,135){.}
\put(242,134){.}
\put(241,133){.}
\put(240,132){.}
\put(239,131){.}
\put(238,130){.}
\put(237,129){.}
\put(236,129){.}
\put(235,128){.}
\put(234,127){.}
\put(233,126){.}
\put(232,125){.}
\put(231,124){.}
\put(230,123){.}
\put(229,123){.}
\put(228,122){.}
\put(227,121){.}
\put(226,120){.}
\put(225,119){.}
\put(224,118){.}
\put(223,117){.}
\put(222,117){.}
\put(221,116){.}
\put(220,115){.}
\put(219,114){.}
\put(218,113){.}
\put(217,113){.}
\put(216,112){.}
\put(215,111){.}
\put(214,110){.}
\put(213,109){.}
\put(212,109){.}
\put(211,108){.}
\put(210,107){.}
\put(209,106){.}
\put(208,105){.}
\put(207,105){.}
\put(206,104){.}
\put(205,103){.}
\put(204,102){.}
\put(203,102){.}
\put(202,101){.}
\put(201,100){.}
\put(200,99){.}
\put(199,99){.}
\put(198,98){.}
\put(197,97){.}
\put(196,96){.}
\put(195,95){.}
\put(194,95){.}
\put(193,94){.}
\put(192,93){.}
\put(191,93){.}
\put(190,92){.}
\put(189,91){.}
\put(188,90){.}
\put(187,90){.}
\put(186,89){.}
\put(185,88){.}
\put(184,87){.}
\put(183,87){.}
\put(182,86){.}
\put(181,85){.}
\put(180,85){.}
\put(179,84){.}
\put(178,83){.}
\put(177,83){.}
\put(176,82){.}
\put(175,81){.}
\put(174,81){.}
\put(173,80){.}
\put(172,79){.}
\put(171,79){.}
\put(170,78){.}
\put(169,77){.}
\put(168,77){.}
\put(167,76){.}
\put(166,75){.}
\put(165,75){.}
\put(164,74){.}
\put(163,73){.}
\put(162,73){.}
\put(161,72){.}
\put(160,71){.}
\put(159,71){.}
\put(158,70){.}
\put(157,70){.}
\put(156,69){.}
\put(155,68){.}
\put(154,68){.}
\put(153,67){.}
\put(152,67){.}
\put(151,66){.}
\put(150,65){.}
\put(149,65){.}
\put(148,64){.}
\put(147,64){.}
\put(146,63){.}
\put(145,62){.}
\put(144,62){.}
\put(143,61){.}
\put(142,61){.}
\put(141,60){.}
\put(140,60){.}
\put(139,59){.}
\put(138,59){.}
\put(137,58){.}
\put(136,58){.}
\put(135,57){.}
\put(134,56){.}
\put(133,56){.}
\put(132,55){.}
\put(131,55){.}
\put(130,54){.}
\put(129,54){.}
\put(128,53){.}
\put(127,53){.}
\put(126,52){.}
\put(125,52){.}
\put(124,52){.}
\put(123,51){.}
\put(122,51){.}
\put(121,50){.}
\put(120,50){.}
\put(119,49){.}
\put(118,49){.}
\put(117,48){.}
\put(116,48){.}
\put(115,47){.}
\put(114,47){.}
\put(113,47){.}
\put(112,46){.}
\put(111,46){.}
\put(110,45){.}
\put(109,45){.}
\put(108,44){.}
\put(107,44){.}
\put(106,44){.}
\put(105,43){.}
\put(104,43){.}
\put(103,43){.}
\put(102,42){.}
\put(101,42){.}
\put(100,41){.}
\put(99,41){.}
\put(98,41){.}
\put(97,40){.}
\put(96,40){.}
\put(95,40){.}
\put(94,39){.}
\put(93,39){.}
\put(92,39){.}
\put(91,38){.}
\put(90,38){.}
\put(89,38){.}
\put(88,37){.}
\put(87,37){.}
\put(86,37){.}
\put(85,37){.}
\put(84,36){.}
\put(83,36){.}
\put(82,36){.}
\put(81,35){.}
\put(80,35){.}
\put(79,35){.}
\put(78,35){.}
\put(77,34){.}
\put(76,34){.}
\put(75,34){.}
\put(74,34){.}
\put(73,33){.}
\put(72,33){.}
\put(71,33){.}
\put(70,33){.}
\put(69,33){.}
\put(68,32){.}
\put(67,32){.}
\put(66,32){.}
\put(65,32){.}
\put(64,31){.}
\put(63,31){.}
\put(62,31){.}
\put(61,31){.}
\put(60,31){.}
\put(59,31){.}
\put(58,30){.}
\put(57,30){.}
\put(56,30){.}
\put(55,30){.}
\put(54,30){.}
\put(53,30){.}
\put(52,29){.}
\put(51,29){.}
\put(50,29){.}
\put(49,29){.}
\put(48,29){.}
\put(47,29){.}
\put(46,29){.}
\put(45,28){.}
\put(44,28){.}
\put(43,28){.}
\put(42,28){.}
\put(41,28){.}
\put(40,28){.}
\put(39,28){.}
\put(38,28){.}
\put(37,28){.}
\put(36,27){.}
\put(35,27){.}
\put(34,27){.}
\put(33,27){.}
\put(32,27){.}
\put(31,27){.}
\put(30,27){.}
\put(29,27){.}
\put(28,27){.}
\put(27,27){.}
\put(26,27){.}
\put(25,27){.}
\put(325,225){.}
\put(324,225){.}
\put(323,225){.}
\put(322,225){.}
\put(321,225){.}
\put(320,225){.}
\put(319,225){.}
\put(318,225){.}
\put(317,225){.}
\put(316,225){.}
\put(315,225){.}
\put(314,225){.}
\put(313,225){.}
\put(312,225){.}
\put(311,225){.}
\put(310,225){.}
\put(309,225){.}
\put(308,225){.}
\put(307,225){.}
\put(306,225){.}
\put(305,225){.}
\put(304,225){.}
\put(303,225){.}
\put(302,225){.}
\put(301,225){.}
\put(300,225){.}
\put(299,225){.}
\put(298,225){.}
\put(297,225){.}
\put(296,225){.}
\put(295,225){.}
\put(294,225){.}
\put(293,225){.}
\put(292,225){.}
\put(291,225){.}
\put(290,225){.}
\put(289,225){.}
\put(288,225){.}
\put(287,225){.}
\put(286,225){.}
\put(285,225){.}
\put(284,225){.}
\put(283,225){.}
\put(282,225){.}
\put(281,225){.}
\put(280,225){.}
\put(279,224){.}
\put(278,224){.}
\put(277,224){.}
\put(276,224){.}
\put(275,224){.}
\put(274,224){.}
\put(273,224){.}
\put(272,224){.}
\put(271,224){.}
\put(270,224){.}
\put(269,224){.}
\put(268,224){.}
\put(267,224){.}
\put(266,224){.}
\put(265,224){.}
\put(264,223){.}
\put(263,223){.}
\put(262,223){.}
\put(261,223){.}
\put(260,223){.}
\put(259,223){.}
\put(258,223){.}
\put(257,223){.}
\put(256,222){.}
\put(255,222){.}
\put(254,222){.}
\put(253,222){.}
\put(252,222){.}
\put(251,222){.}
\put(250,222){.}
\put(249,221){.}
\put(248,221){.}
\put(247,221){.}
\put(246,221){.}
\put(245,221){.}
\put(244,220){.}
\put(243,220){.}
\put(242,220){.}
\put(241,220){.}
\put(240,219){.}
\put(239,219){.}
\put(238,219){.}
\put(237,219){.}
\put(236,218){.}
\put(235,218){.}
\put(234,218){.}
\put(233,217){.}
\put(232,217){.}
\put(231,217){.}
\put(230,216){.}
\put(229,216){.}
\put(228,216){.}
\put(227,215){.}
\put(226,215){.}
\put(225,214){.}
\put(224,214){.}
\put(223,214){.}
\put(222,213){.}
\put(221,213){.}
\put(220,212){.}
\put(219,212){.}
\put(218,211){.}
\put(217,211){.}
\put(216,210){.}
\put(215,210){.}
\put(214,209){.}
\put(213,209){.}
\put(212,208){.}
\put(211,208){.}
\put(210,207){.}
\put(209,207){.}
\put(208,206){.}
\put(207,205){.}
\put(206,205){.}
\put(205,204){.}
\put(204,203){.}
\put(203,203){.}
\put(202,202){.}
\put(201,202){.}
\put(200,201){.}
\put(199,200){.}
\put(198,199){.}
\put(197,199){.}
\put(196,198){.}
\put(195,197){.}
\put(194,196){.}
\put(193,196){.}
\put(192,195){.}
\put(191,194){.}
\put(190,193){.}
\put(189,192){.}
\put(188,192){.}
\put(187,191){.}
\put(186,190){.}
\put(185,189){.}
\put(184,188){.}
\put(183,187){.}
\put(182,186){.}
\put(181,185){.}
\put(180,184){.}
\put(179,183){.}
\put(178,182){.}
\put(177,181){.}
\put(176,180){.}
\put(175,179){.}
\put(174,178){.}
\put(173,177){.}
\put(172,176){.}
\put(171,175){.}
\put(170,174){.}
\put(169,173){.}
\put(168,172){.}
\put(167,171){.}
\put(166,170){.}
\put(165,168){.}
\put(164,167){.}
\put(163,166){.}
\put(162,165){.}
\put(161,164){.}
\put(160,163){.}
\put(159,162){.}
\put(158,160){.}
\put(157,159){.}
\put(156,158){.}
\put(155,157){.}
\put(154,155){.}
\put(153,154){.}
\put(152,153){.}
\put(151,152){.}
\put(150,150){.}
\put(149,149){.}
\put(148,148){.}
\put(147,147){.}
\put(146,145){.}
\put(145,144){.}
\put(144,143){.}
\put(143,141){.}
\put(142,140){.}
\put(141,139){.}
\put(140,137){.}
\put(139,136){.}
\put(138,135){.}
\put(137,133){.}
\put(136,132){.}
\put(135,131){.}
\put(134,129){.}
\put(133,128){.}
\put(132,127){.}
\put(131,125){.}
\put(130,124){.}
\put(129,123){.}
\put(128,121){.}
\put(127,120){.}
\put(126,119){.}
\put(125,117){.}
\put(124,116){.}
\put(123,115){.}
\put(122,113){.}
\put(121,112){.}
\put(120,111){.}
\put(119,109){.}
\put(118,108){.}
\put(117,107){.}
\put(116,105){.}
\put(115,104){.}
\put(114,103){.}
\put(113,102){.}
\put(112,100){.}
\put(111,99){.}
\put(110,98){.}
\put(109,96){.}
\put(108,95){.}
\put(107,94){.}
\put(106,93){.}
\put(105,91){.}
\put(104,90){.}
\put(103,89){.}
\put(102,88){.}
\put(101,87){.}
\put(100,85){.}
\put(99,84){.}
\put(98,83){.}
\put(97,82){.}
\put(96,81){.}
\put(95,79){.}
\put(94,78){.}
\put(93,77){.}
\put(92,76){.}
\put(91,75){.}
\put(90,74){.}
\put(89,73){.}
\put(88,72){.}
\put(87,71){.}
\put(86,70){.}
\put(85,69){.}
\put(84,68){.}
\put(83,67){.}
\put(82,66){.}
\put(81,64){.}
\put(80,64){.}
\put(79,63){.}
\put(78,62){.}
\put(77,61){.}
\put(76,60){.}
\put(75,59){.}
\put(74,58){.}
\put(73,57){.}
\put(72,56){.}
\put(71,55){.}
\put(70,55){.}
\put(69,54){.}
\put(68,53){.}
\put(67,52){.}
\put(66,52){.}
\put(65,51){.}
\put(64,50){.}
\put(63,49){.}
\put(62,49){.}
\put(61,48){.}
\put(60,47){.}
\put(59,47){.}
\put(58,46){.}
\put(57,45){.}
\put(56,45){.}
\put(55,44){.}
\put(54,43){.}
\put(53,43){.}
\put(52,42){.}
\put(51,42){.}
\put(50,41){.}
\put(49,41){.}
\put(48,40){.}
\put(47,40){.}
\put(46,39){.}
\put(45,39){.}
\put(44,38){.}
\put(43,38){.}
\put(42,37){.}
\put(41,37){.}
\put(40,36){.}
\put(39,36){.}
\put(38,36){.}
\put(37,35){.}
\put(36,35){.}
\put(35,34){.}
\put(34,34){.}
\put(33,34){.}
\put(32,33){.}
\put(31,33){.}
\put(30,33){.}
\put(29,32){.}
\put(28,32){.}
\put(27,32){.}
\put(26,32){.}
\put(25,31){.}
\put(325,225){.}
\put(319,217){.}
\put(318,216){.}
\put(317,214){.}
\put(316,213){.}
\put(315,212){.}
\put(309,204){.}
\put(308,203){.}
\put(307,201){.}
\put(306,200){.}
\put(305,199){.}
\put(299,192){.}
\put(298,191){.}
\put(297,189){.}
\put(296,188){.}
\put(295,187){.}
\put(289,180){.}
\put(288,179){.}
\put(287,178){.}
\put(286,177){.}
\put(285,176){.}
\put(279,169){.}
\put(278,168){.}
\put(277,167){.}
\put(276,166){.}
\put(275,165){.}
\put(269,159){.}
\put(268,158){.}
\put(267,157){.}
\put(266,156){.}
\put(265,155){.}
\put(259,149){.}
\put(258,148){.}
\put(257,147){.}
\put(256,146){.}
\put(255,146){.}
\put(249,140){.}
\put(248,139){.}
\put(247,138){.}
\put(246,137){.}
\put(245,136){.}
\put(239,131){.}
\put(238,130){.}
\put(237,129){.}
\put(236,128){.}
\put(235,127){.}
\put(229,122){.}
\put(228,121){.}
\put(227,121){.}
\put(226,120){.}
\put(225,119){.}
\put(219,114){.}
\put(218,113){.}
\put(217,112){.}
\put(216,111){.}
\put(215,111){.}
\put(209,106){.}
\put(208,105){.}
\put(207,104){.}
\put(206,103){.}
\put(205,103){.}
\put(199,98){.}
\put(198,97){.}
\put(197,96){.}
\put(196,96){.}
\put(195,95){.}
\put(189,90){.}
\put(188,90){.}
\put(187,89){.}
\put(186,88){.}
\put(185,87){.}
\put(179,83){.}
\put(178,82){.}
\put(177,82){.}
\put(176,81){.}
\put(175,80){.}
\put(169,76){.}
\put(168,76){.}
\put(167,75){.}
\put(166,74){.}
\put(165,73){.}
\put(159,70){.}
\put(158,69){.}
\put(157,68){.}
\put(156,68){.}
\put(155,67){.}
\put(149,63){.}
\put(148,63){.}
\put(147,62){.}
\put(146,62){.}
\put(145,61){.}
\put(139,58){.}
\put(138,57){.}
\put(137,56){.}
\put(136,56){.}
\put(135,55){.}
\put(129,52){.}
\put(128,52){.}
\put(127,51){.}
\put(126,51){.}
\put(125,50){.}
\put(119,47){.}
\put(118,47){.}
\put(117,46){.}
\put(116,46){.}
\put(115,46){.}
\put(109,43){.}
\put(108,43){.}
\put(107,42){.}
\put(106,42){.}
\put(105,41){.}
\put(99,39){.}
\put(98,39){.}
\put(97,39){.}
\put(96,38){.}
\put(95,38){.}
\put(89,36){.}
\put(88,36){.}
\put(87,35){.}
\put(86,35){.}
\put(85,35){.}
\put(79,33){.}
\put(78,33){.}
\put(77,33){.}
\put(76,33){.}
\put(75,32){.}
\put(69,31){.}
\put(68,31){.}
\put(67,31){.}
\put(66,30){.}
\put(65,30){.}
\put(59,29){.}
\put(58,29){.}
\put(57,29){.}
\put(56,29){.}
\put(55,29){.}
\put(49,28){.}
\put(48,28){.}
\put(47,28){.}
\put(46,28){.}
\put(45,27){.}
\put(39,27){.}
\put(38,27){.}
\put(37,27){.}
\put(36,27){.}
\put(35,27){.}
\put(29,26){.}
\put(28,26){.}
\put(27,26){.}
\put(26,26){.}
\put(25,26){.}
\put(325,225){.}
\put(319,225){.}
\put(318,225){.}
\put(317,225){.}
\put(316,225){.}
\put(315,225){.}
\put(309,225){.}
\put(308,225){.}
\put(307,225){.}
\put(306,225){.}
\put(305,225){.}
\put(299,225){.}
\put(298,225){.}
\put(297,225){.}
\put(296,225){.}
\put(295,225){.}
\put(289,225){.}
\put(288,225){.}
\put(287,225){.}
\put(286,225){.}
\put(285,225){.}
\put(279,224){.}
\put(278,224){.}
\put(277,224){.}
\put(276,224){.}
\put(275,224){.}
\put(269,224){.}
\put(268,224){.}
\put(267,224){.}
\put(266,224){.}
\put(265,223){.}
\put(259,223){.}
\put(258,223){.}
\put(257,223){.}
\put(256,222){.}
\put(255,222){.}
\put(249,221){.}
\put(248,221){.}
\put(247,221){.}
\put(246,221){.}
\put(245,220){.}
\put(239,219){.}
\put(238,218){.}
\put(237,218){.}
\put(236,218){.}
\put(235,218){.}
\put(229,215){.}
\put(228,215){.}
\put(227,215){.}
\put(226,214){.}
\put(225,214){.}
\put(219,211){.}
\put(218,211){.}
\put(217,210){.}
\put(216,210){.}
\put(215,209){.}
\put(209,205){.}
\put(208,205){.}
\put(207,204){.}
\put(206,204){.}
\put(205,203){.}
\put(199,199){.}
\put(198,198){.}
\put(197,197){.}
\put(196,196){.}
\put(195,195){.}
\put(189,190){.}
\put(188,189){.}
\put(187,188){.}
\put(186,187){.}
\put(185,186){.}
\put(179,180){.}
\put(178,179){.}
\put(177,178){.}
\put(176,177){.}
\put(175,176){.}
\put(169,169){.}
\put(168,168){.}
\put(167,167){.}
\put(166,166){.}
\put(165,165){.}
\put(159,157){.}
\put(158,156){.}
\put(157,154){.}
\put(156,153){.}
\put(155,152){.}
\put(149,144){.}
\put(148,142){.}
\put(147,141){.}
\put(146,140){.}
\put(145,138){.}
\put(139,130){.}
\put(138,129){.}
\put(137,127){.}
\put(136,126){.}
\put(135,124){.}
\put(129,116){.}
\put(128,115){.}
\put(127,113){.}
\put(126,112){.}
\put(125,110){.}
\put(119,102){.}
\put(118,101){.}
\put(117,99){.}
\put(116,98){.}
\put(115,97){.}
\put(109,89){.}
\put(108,87){.}
\put(107,86){.}
\put(106,85){.}
\put(105,84){.}
\put(99,76){.}
\put(98,75){.}
\put(97,74){.}
\put(96,73){.}
\put(95,72){.}
\put(89,65){.}
\put(88,64){.}
\put(87,63){.}
\put(86,62){.}
\put(85,61){.}
\put(79,56){.}
\put(78,55){.}
\put(77,54){.}
\put(76,53){.}
\put(75,52){.}
\put(69,47){.}
\put(68,47){.}
\put(67,46){.}
\put(66,45){.}
\put(65,45){.}
\put(59,41){.}
\put(58,40){.}
\put(57,40){.}
\put(56,39){.}
\put(55,39){.}
\put(49,36){.}
\put(48,36){.}
\put(47,35){.}
\put(46,35){.}
\put(45,34){.}
\put(39,32){.}
\put(38,32){.}
\put(37,32){.}
\put(36,31){.}
\put(35,31){.}
\put(29,30){.}
\put(28,29){.}
\put(27,29){.}
\put(26,29){.}
\put(25,29){.}
\put(25,25){\vector(0,1){210}}
\put(25,25){\vector(1,0){310}}
\put(325,23){\line(0,1){4}}
\put(22,225){\line(1,0){4}}
\put(325,18){$z$}
\put(13,225){$f$}
\put(340,18){$\rho$}
\put(18,240){$distributions$}
\put(0,0){\line(1,0){400}}
\put(0,0){\line(0,1){300}}
\put(0,300){\line(1,0){400}}
\put(400,0){\line(0,1){300}}
\end{picture}

\begin{center}

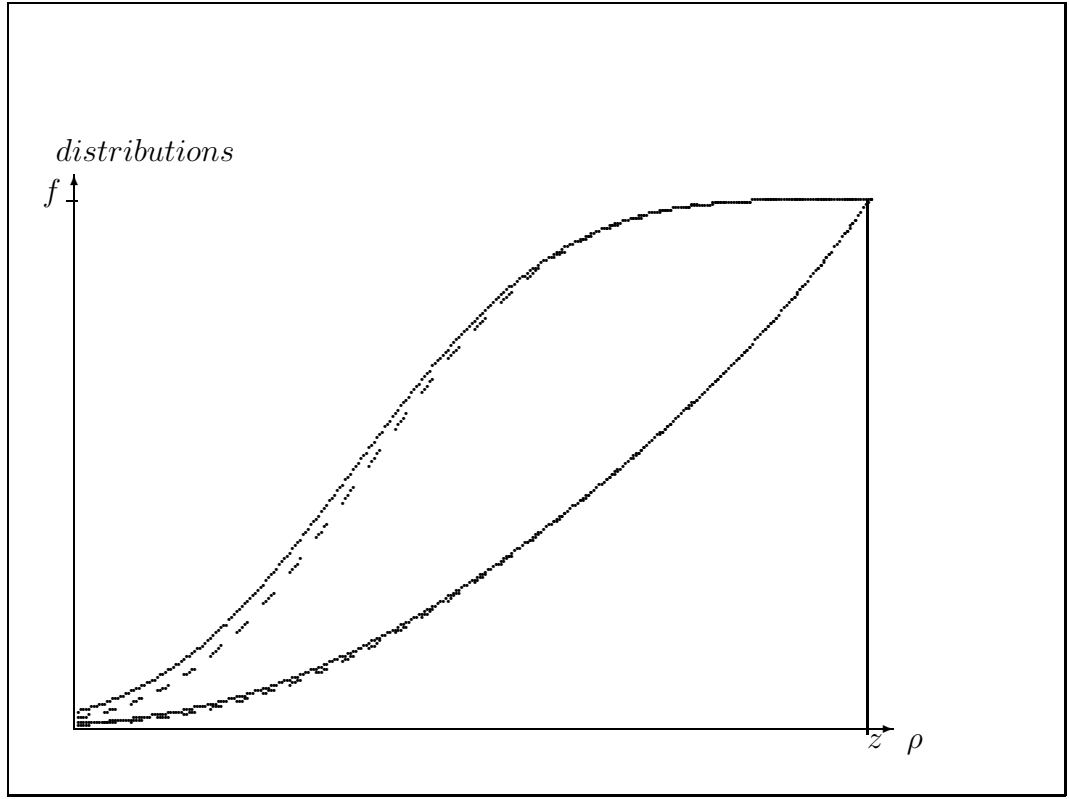
{\it Fig.3}

\end{center}

{\it  One can see two curves which are going from $\rho = z$ to the small
sizes $\rho$. The lower curve corresponds to the real spectrum calculated
with account of heterogeneous centers exhaustion. The upper curve corresponds
to the condensation without exhaustion of heterogeneous centers which
is the worst situation where there is converging force
due to the centers exhaustion.

Concrete situation drawn here corresponds to $b=2$ after renormalization
(the values of parameters
$f$ and $\Gamma$ can be canceled by appropriate renormalization). The
value of $z$ here equals to $3 \Delta z / 2$.

The solid lines correspond to the precise numerical solution. The dashed
lines correspond to  application of  monodisperse approximation.
One can not separate the numerical solutions from the approximate ones
except the slight deviation in the region of small $\rho$. As far as all
(precise and approximate) solutions will go to zero there will be no deviations
for $z \gg \Delta z$ (i.e. we stop at the worst moment).  }

\pagebreak

\begin{picture}(300,300)
\put(26,124){.}
\put(27,131){.}
\put(28,139){.}
\put(30,146){.}
\put(31,152){.}
\put(32,159){.}
\put(34,165){.}
\put(35,171){.}
\put(36,177){.}
\put(38,182){.}
\put(39,187){.}
\put(40,192){.}
\put(42,197){.}
\put(43,202){.}
\put(44,206){.}
\put(46,210){.}
\put(47,214){.}
\put(48,217){.}
\put(50,221){.}
\put(51,224){.}
\put(52,227){.}
\put(54,230){.}
\put(55,233){.}
\put(56,235){.}
\put(58,237){.}
\put(59,239){.}
\put(60,241){.}
\put(62,243){.}
\put(63,245){.}
\put(64,246){.}
\put(66,247){.}
\put(67,248){.}
\put(68,249){.}
\put(70,250){.}
\put(71,250){.}
\put(72,251){.}
\put(74,251){.}
\put(75,251){.}
\put(76,251){.}
\put(78,251){.}
\put(79,251){.}
\put(80,251){.}
\put(82,250){.}
\put(83,250){.}
\put(84,249){.}
\put(86,248){.}
\put(87,247){.}
\put(88,246){.}
\put(90,245){.}
\put(91,244){.}
\put(92,242){.}
\put(94,241){.}
\put(95,239){.}
\put(96,238){.}
\put(98,236){.}
\put(99,234){.}
\put(100,233){.}
\put(102,231){.}
\put(103,229){.}
\put(104,227){.}
\put(106,225){.}
\put(107,222){.}
\put(108,220){.}
\put(110,218){.}
\put(111,216){.}
\put(112,213){.}
\put(114,211){.}
\put(115,209){.}
\put(116,206){.}
\put(118,203){.}
\put(119,201){.}
\put(120,198){.}
\put(122,196){.}
\put(123,193){.}
\put(124,190){.}
\put(126,188){.}
\put(127,185){.}
\put(128,182){.}
\put(130,180){.}
\put(131,177){.}
\put(132,174){.}
\put(134,171){.}
\put(135,169){.}
\put(136,166){.}
\put(138,163){.}
\put(139,161){.}
\put(140,158){.}
\put(142,155){.}
\put(143,152){.}
\put(144,150){.}
\put(146,147){.}
\put(147,144){.}
\put(148,142){.}
\put(150,139){.}
\put(151,137){.}
\put(152,134){.}
\put(154,131){.}
\put(155,129){.}
\put(156,126){.}
\put(158,124){.}
\put(159,122){.}
\put(160,119){.}
\put(162,117){.}
\put(163,114){.}
\put(164,112){.}
\put(166,110){.}
\put(167,108){.}
\put(168,106){.}
\put(170,103){.}
\put(171,101){.}
\put(172,99){.}
\put(174,97){.}
\put(175,95){.}
\put(176,93){.}
\put(178,91){.}
\put(179,90){.}
\put(180,88){.}
\put(182,86){.}
\put(183,84){.}
\put(184,83){.}
\put(186,81){.}
\put(187,79){.}
\put(188,78){.}
\put(190,76){.}
\put(191,75){.}
\put(192,73){.}
\put(194,72){.}
\put(195,71){.}
\put(196,69){.}
\put(198,68){.}
\put(199,67){.}
\put(200,66){.}
\put(202,65){.}
\put(203,63){.}
\put(204,62){.}
\put(206,61){.}
\put(207,60){.}
\put(208,59){.}
\put(210,58){.}
\put(211,57){.}
\put(212,56){.}
\put(214,55){.}
\put(215,55){.}
\put(216,54){.}
\put(218,53){.}
\put(219,52){.}
\put(220,51){.}
\put(222,51){.}
\put(223,50){.}
\put(224,49){.}
\put(25,25){\vector(0,1){210}}
\put(25,25){\vector(1,0){210}}
\put(93,23){\line(0,1){4}}
\put(22,225){\line(1,0){4}}
\put(93,18){1}
\put(3,225){$0.01$}
\put(235,18){$b$}
\put(18,240){$error$}
\put(0,0){\line(1,0){300}}
\put(0,0){\line(0,1){300}}
\put(0,300){\line(1,0){300}}
\put(300,0){\line(0,1){300}}
\end{picture}

\begin{center}

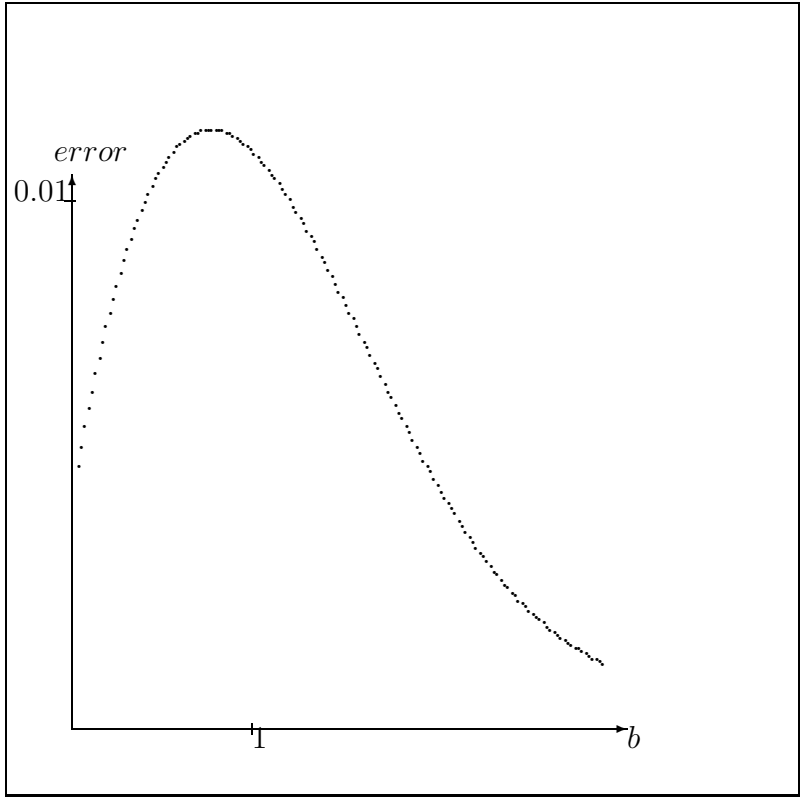
{\it Fig.4}

\end{center}

{\it  The relative error of approximate solution for the nucleation on
the similar heterogeneous centers. Here the values $f$ and $\Gamma$ can
be canceled after renormalization and there remains only one parameter
$b$ which is the argument of the function drawn here. It is clear that
this function is small. All asymptotes can be checked analytically (see
\cite{Ierarchy}). }

\pagebreak

\begin{picture}(400,300)
\put(325,225){\line(0,-1){200}}
\put(325,225){.}
\put(325,225){.}
\put(325,225){.}
\put(324,223){.}
\put(324,225){.}
\put(324,225){.}
\put(323,219){.}
\put(323,224){.}
\put(323,225){.}
\put(322,216){.}
\put(322,223){.}
\put(322,225){.}
\put(321,212){.}
\put(321,222){.}
\put(321,225){.}
\put(320,209){.}
\put(320,221){.}
\put(320,225){.}
\put(319,205){.}
\put(319,220){.}
\put(319,225){.}
\put(318,202){.}
\put(318,219){.}
\put(318,225){.}
\put(317,199){.}
\put(317,218){.}
\put(317,225){.}
\put(316,195){.}
\put(316,217){.}
\put(316,225){.}
\put(315,192){.}
\put(315,216){.}
\put(315,225){.}
\put(314,189){.}
\put(314,215){.}
\put(314,225){.}
\put(313,186){.}
\put(313,214){.}
\put(313,225){.}
\put(312,183){.}
\put(312,214){.}
\put(312,225){.}
\put(311,180){.}
\put(311,213){.}
\put(311,225){.}
\put(310,177){.}
\put(310,212){.}
\put(310,225){.}
\put(309,174){.}
\put(309,211){.}
\put(309,225){.}
\put(308,171){.}
\put(308,210){.}
\put(308,225){.}
\put(307,169){.}
\put(307,209){.}
\put(307,225){.}
\put(306,166){.}
\put(306,208){.}
\put(306,225){.}
\put(305,163){.}
\put(305,207){.}
\put(305,225){.}
\put(304,161){.}
\put(304,206){.}
\put(304,225){.}
\put(303,158){.}
\put(303,206){.}
\put(303,225){.}
\put(302,156){.}
\put(302,205){.}
\put(302,225){.}
\put(301,153){.}
\put(301,204){.}
\put(301,225){.}
\put(300,151){.}
\put(300,203){.}
\put(300,225){.}
\put(299,148){.}
\put(299,202){.}
\put(299,225){.}
\put(298,146){.}
\put(298,201){.}
\put(298,225){.}
\put(297,144){.}
\put(297,200){.}
\put(297,225){.}
\put(296,142){.}
\put(296,200){.}
\put(296,225){.}
\put(295,139){.}
\put(295,199){.}
\put(295,225){.}
\put(294,137){.}
\put(294,198){.}
\put(294,225){.}
\put(293,135){.}
\put(293,197){.}
\put(293,224){.}
\put(292,133){.}
\put(292,196){.}
\put(292,224){.}
\put(291,131){.}
\put(291,195){.}
\put(291,224){.}
\put(290,129){.}
\put(290,194){.}
\put(290,224){.}
\put(289,127){.}
\put(289,193){.}
\put(289,224){.}
\put(288,125){.}
\put(288,193){.}
\put(288,224){.}
\put(287,123){.}
\put(287,192){.}
\put(287,224){.}
\put(286,121){.}
\put(286,191){.}
\put(286,224){.}
\put(285,119){.}
\put(285,190){.}
\put(285,224){.}
\put(284,117){.}
\put(284,189){.}
\put(284,224){.}
\put(283,116){.}
\put(283,188){.}
\put(283,223){.}
\put(282,114){.}
\put(282,187){.}
\put(282,223){.}
\put(281,112){.}
\put(281,186){.}
\put(281,223){.}
\put(280,110){.}
\put(280,185){.}
\put(280,223){.}
\put(279,109){.}
\put(279,184){.}
\put(279,223){.}
\put(278,107){.}
\put(278,184){.}
\put(278,223){.}
\put(277,106){.}
\put(277,183){.}
\put(277,222){.}
\put(276,104){.}
\put(276,182){.}
\put(276,222){.}
\put(275,102){.}
\put(275,181){.}
\put(275,222){.}
\put(274,101){.}
\put(274,180){.}
\put(274,222){.}
\put(273,99){.}
\put(273,179){.}
\put(273,221){.}
\put(272,98){.}
\put(272,178){.}
\put(272,221){.}
\put(271,96){.}
\put(271,177){.}
\put(271,221){.}
\put(270,95){.}
\put(270,176){.}
\put(270,220){.}
\put(269,94){.}
\put(269,175){.}
\put(269,220){.}
\put(268,92){.}
\put(268,174){.}
\put(268,220){.}
\put(267,91){.}
\put(267,173){.}
\put(267,219){.}
\put(266,90){.}
\put(266,172){.}
\put(266,219){.}
\put(265,88){.}
\put(265,171){.}
\put(265,219){.}
\put(264,87){.}
\put(264,170){.}
\put(264,218){.}
\put(263,86){.}
\put(263,169){.}
\put(263,218){.}
\put(262,84){.}
\put(262,168){.}
\put(262,217){.}
\put(261,83){.}
\put(261,167){.}
\put(261,217){.}
\put(260,82){.}
\put(260,166){.}
\put(260,216){.}
\put(259,81){.}
\put(259,165){.}
\put(259,216){.}
\put(258,80){.}
\put(258,164){.}
\put(258,215){.}
\put(257,79){.}
\put(257,163){.}
\put(257,215){.}
\put(256,78){.}
\put(256,162){.}
\put(256,214){.}
\put(255,76){.}
\put(255,161){.}
\put(255,214){.}
\put(254,75){.}
\put(254,160){.}
\put(254,213){.}
\put(253,74){.}
\put(253,159){.}
\put(253,212){.}
\put(252,73){.}
\put(252,158){.}
\put(252,212){.}
\put(251,72){.}
\put(251,157){.}
\put(251,211){.}
\put(250,71){.}
\put(250,156){.}
\put(250,210){.}
\put(249,70){.}
\put(249,155){.}
\put(249,210){.}
\put(248,69){.}
\put(248,154){.}
\put(248,209){.}
\put(247,68){.}
\put(247,153){.}
\put(247,208){.}
\put(246,67){.}
\put(246,152){.}
\put(246,207){.}
\put(245,66){.}
\put(245,150){.}
\put(245,206){.}
\put(244,66){.}
\put(244,149){.}
\put(244,205){.}
\put(243,65){.}
\put(243,148){.}
\put(243,205){.}
\put(242,64){.}
\put(242,147){.}
\put(242,204){.}
\put(241,63){.}
\put(241,146){.}
\put(241,203){.}
\put(240,62){.}
\put(240,145){.}
\put(240,202){.}
\put(239,61){.}
\put(239,144){.}
\put(239,201){.}
\put(238,60){.}
\put(238,142){.}
\put(238,200){.}
\put(237,60){.}
\put(237,141){.}
\put(237,199){.}
\put(236,59){.}
\put(236,140){.}
\put(236,198){.}
\put(235,58){.}
\put(235,139){.}
\put(235,197){.}
\put(234,57){.}
\put(234,137){.}
\put(234,195){.}
\put(233,57){.}
\put(233,136){.}
\put(233,194){.}
\put(232,56){.}
\put(232,135){.}
\put(232,193){.}
\put(231,55){.}
\put(231,134){.}
\put(231,192){.}
\put(230,55){.}
\put(230,133){.}
\put(230,190){.}
\put(229,54){.}
\put(229,131){.}
\put(229,189){.}
\put(228,53){.}
\put(228,130){.}
\put(228,188){.}
\put(227,53){.}
\put(227,129){.}
\put(227,187){.}
\put(226,52){.}
\put(226,128){.}
\put(226,185){.}
\put(225,51){.}
\put(225,126){.}
\put(225,184){.}
\put(224,51){.}
\put(224,125){.}
\put(224,182){.}
\put(223,50){.}
\put(223,124){.}
\put(223,181){.}
\put(222,49){.}
\put(222,122){.}
\put(222,180){.}
\put(221,49){.}
\put(221,121){.}
\put(221,178){.}
\put(220,48){.}
\put(220,120){.}
\put(220,177){.}
\put(219,48){.}
\put(219,119){.}
\put(219,175){.}
\put(218,47){.}
\put(218,117){.}
\put(218,174){.}
\put(217,47){.}
\put(217,116){.}
\put(217,172){.}
\put(216,46){.}
\put(216,115){.}
\put(216,170){.}
\put(215,46){.}
\put(215,113){.}
\put(215,169){.}
\put(214,45){.}
\put(214,112){.}
\put(214,167){.}
\put(213,45){.}
\put(213,111){.}
\put(213,165){.}
\put(212,44){.}
\put(212,110){.}
\put(212,164){.}
\put(211,44){.}
\put(211,108){.}
\put(211,162){.}
\put(210,43){.}
\put(210,107){.}
\put(210,160){.}
\put(209,43){.}
\put(209,106){.}
\put(209,159){.}
\put(208,42){.}
\put(208,104){.}
\put(208,157){.}
\put(207,42){.}
\put(207,103){.}
\put(207,155){.}
\put(206,41){.}
\put(206,102){.}
\put(206,153){.}
\put(205,41){.}
\put(205,100){.}
\put(205,151){.}
\put(204,41){.}
\put(204,99){.}
\put(204,150){.}
\put(203,40){.}
\put(203,98){.}
\put(203,148){.}
\put(202,40){.}
\put(202,96){.}
\put(202,146){.}
\put(201,39){.}
\put(201,95){.}
\put(201,144){.}
\put(200,39){.}
\put(200,94){.}
\put(200,142){.}
\put(199,39){.}
\put(199,93){.}
\put(199,140){.}
\put(198,38){.}
\put(198,91){.}
\put(198,138){.}
\put(197,38){.}
\put(197,90){.}
\put(197,137){.}
\put(196,37){.}
\put(196,89){.}
\put(196,135){.}
\put(195,37){.}
\put(195,87){.}
\put(195,133){.}
\put(194,37){.}
\put(194,86){.}
\put(194,131){.}
\put(193,36){.}
\put(193,85){.}
\put(193,129){.}
\put(192,36){.}
\put(192,84){.}
\put(192,127){.}
\put(191,36){.}
\put(191,82){.}
\put(191,125){.}
\put(190,36){.}
\put(190,81){.}
\put(190,123){.}
\put(189,35){.}
\put(189,80){.}
\put(189,121){.}
\put(188,35){.}
\put(188,79){.}
\put(188,119){.}
\put(187,35){.}
\put(187,77){.}
\put(187,117){.}
\put(186,34){.}
\put(186,76){.}
\put(186,115){.}
\put(185,34){.}
\put(185,75){.}
\put(185,113){.}
\put(184,34){.}
\put(184,74){.}
\put(184,111){.}
\put(183,34){.}
\put(183,73){.}
\put(183,110){.}
\put(182,33){.}
\put(182,72){.}
\put(182,108){.}
\put(181,33){.}
\put(181,70){.}
\put(181,106){.}
\put(180,33){.}
\put(180,69){.}
\put(180,104){.}
\put(179,33){.}
\put(179,68){.}
\put(179,102){.}
\put(178,32){.}
\put(178,67){.}
\put(178,100){.}
\put(177,32){.}
\put(177,66){.}
\put(177,98){.}
\put(176,32){.}
\put(176,65){.}
\put(176,97){.}
\put(175,32){.}
\put(175,64){.}
\put(175,95){.}
\put(174,31){.}
\put(174,63){.}
\put(174,93){.}
\put(173,31){.}
\put(173,62){.}
\put(173,91){.}
\put(172,31){.}
\put(172,61){.}
\put(172,89){.}
\put(171,31){.}
\put(171,60){.}
\put(171,87){.}
\put(170,31){.}
\put(170,59){.}
\put(170,86){.}
\put(169,30){.}
\put(169,57){.}
\put(169,84){.}
\put(168,30){.}
\put(168,57){.}
\put(168,82){.}
\put(167,30){.}
\put(167,56){.}
\put(167,81){.}
\put(166,30){.}
\put(166,55){.}
\put(166,79){.}
\put(165,30){.}
\put(165,54){.}
\put(165,78){.}
\put(164,30){.}
\put(164,53){.}
\put(164,76){.}
\put(163,29){.}
\put(163,52){.}
\put(163,74){.}
\put(162,29){.}
\put(162,51){.}
\put(162,73){.}
\put(161,29){.}
\put(161,50){.}
\put(161,71){.}
\put(160,29){.}
\put(160,49){.}
\put(160,70){.}
\put(159,29){.}
\put(159,49){.}
\put(159,68){.}
\put(158,29){.}
\put(158,48){.}
\put(158,67){.}
\put(157,29){.}
\put(157,47){.}
\put(157,65){.}
\put(156,28){.}
\put(156,46){.}
\put(156,64){.}
\put(155,28){.}
\put(155,45){.}
\put(155,63){.}
\put(154,28){.}
\put(154,45){.}
\put(154,61){.}
\put(153,28){.}
\put(153,44){.}
\put(153,60){.}
\put(152,28){.}
\put(152,43){.}
\put(152,59){.}
\put(151,28){.}
\put(151,42){.}
\put(151,57){.}
\put(150,28){.}
\put(150,42){.}
\put(150,56){.}
\put(149,28){.}
\put(149,41){.}
\put(149,55){.}
\put(148,27){.}
\put(148,40){.}
\put(148,54){.}
\put(147,27){.}
\put(147,40){.}
\put(147,53){.}
\put(146,27){.}
\put(146,39){.}
\put(146,52){.}
\put(145,27){.}
\put(145,39){.}
\put(145,50){.}
\put(144,27){.}
\put(144,38){.}
\put(144,49){.}
\put(143,27){.}
\put(143,38){.}
\put(143,48){.}
\put(142,27){.}
\put(142,37){.}
\put(142,47){.}
\put(141,27){.}
\put(141,36){.}
\put(141,46){.}
\put(140,27){.}
\put(140,36){.}
\put(140,46){.}
\put(139,27){.}
\put(139,35){.}
\put(139,45){.}
\put(138,27){.}
\put(138,35){.}
\put(138,44){.}
\put(137,26){.}
\put(137,35){.}
\put(137,43){.}
\put(136,26){.}
\put(136,34){.}
\put(136,42){.}
\put(135,26){.}
\put(135,34){.}
\put(135,41){.}
\put(134,26){.}
\put(134,33){.}
\put(134,40){.}
\put(133,26){.}
\put(133,33){.}
\put(133,40){.}
\put(132,26){.}
\put(132,33){.}
\put(132,39){.}
\put(131,26){.}
\put(131,32){.}
\put(131,38){.}
\put(130,26){.}
\put(130,32){.}
\put(130,38){.}
\put(129,26){.}
\put(129,31){.}
\put(129,37){.}
\put(128,26){.}
\put(128,31){.}
\put(128,36){.}
\put(127,26){.}
\put(127,31){.}
\put(127,36){.}
\put(126,26){.}
\put(126,30){.}
\put(126,35){.}
\put(125,26){.}
\put(125,30){.}
\put(125,35){.}
\put(124,26){.}
\put(124,30){.}
\put(124,34){.}
\put(123,26){.}
\put(123,30){.}
\put(123,34){.}
\put(122,26){.}
\put(122,29){.}
\put(122,33){.}
\put(121,26){.}
\put(121,29){.}
\put(121,33){.}
\put(120,26){.}
\put(120,29){.}
\put(120,32){.}
\put(119,26){.}
\put(119,29){.}
\put(119,32){.}
\put(118,26){.}
\put(118,28){.}
\put(118,32){.}
\put(117,25){.}
\put(117,28){.}
\put(117,31){.}
\put(116,25){.}
\put(116,28){.}
\put(116,31){.}
\put(115,25){.}
\put(115,28){.}
\put(115,30){.}
\put(114,25){.}
\put(114,28){.}
\put(114,30){.}
\put(113,25){.}
\put(113,28){.}
\put(113,30){.}
\put(112,25){.}
\put(112,27){.}
\put(112,29){.}
\put(111,25){.}
\put(111,27){.}
\put(111,29){.}
\put(110,25){.}
\put(110,27){.}
\put(110,29){.}
\put(109,25){.}
\put(109,27){.}
\put(109,29){.}
\put(108,25){.}
\put(108,27){.}
\put(108,28){.}
\put(107,25){.}
\put(107,27){.}
\put(107,28){.}
\put(106,25){.}
\put(106,27){.}
\put(106,28){.}
\put(105,25){.}
\put(105,26){.}
\put(105,28){.}
\put(104,25){.}
\put(104,26){.}
\put(104,28){.}
\put(103,25){.}
\put(103,26){.}
\put(103,27){.}
\put(102,25){.}
\put(102,26){.}
\put(102,27){.}
\put(101,25){.}
\put(101,26){.}
\put(101,27){.}
\put(100,25){.}
\put(100,26){.}
\put(100,27){.}
\put(99,25){.}
\put(99,26){.}
\put(99,27){.}
\put(98,25){.}
\put(98,26){.}
\put(98,27){.}
\put(97,25){.}
\put(97,26){.}
\put(97,27){.}
\put(96,25){.}
\put(96,26){.}
\put(96,26){.}
\put(95,25){.}
\put(95,26){.}
\put(95,26){.}
\put(94,25){.}
\put(94,26){.}
\put(94,26){.}
\put(93,25){.}
\put(93,26){.}
\put(93,26){.}
\put(92,25){.}
\put(92,26){.}
\put(92,26){.}
\put(91,25){.}
\put(91,26){.}
\put(91,26){.}
\put(90,25){.}
\put(90,25){.}
\put(90,26){.}
\put(89,25){.}
\put(89,25){.}
\put(89,26){.}
\put(88,25){.}
\put(88,25){.}
\put(88,26){.}
\put(87,25){.}
\put(87,25){.}
\put(87,26){.}
\put(86,25){.}
\put(86,25){.}
\put(86,26){.}
\put(85,25){.}
\put(85,25){.}
\put(85,26){.}
\put(84,25){.}
\put(84,25){.}
\put(84,26){.}
\put(83,25){.}
\put(83,25){.}
\put(83,26){.}
\put(82,25){.}
\put(82,25){.}
\put(82,25){.}
\put(81,25){.}
\put(81,25){.}
\put(81,25){.}
\put(80,25){.}
\put(80,25){.}
\put(80,25){.}
\put(79,25){.}
\put(79,25){.}
\put(79,25){.}
\put(78,25){.}
\put(78,25){.}
\put(78,25){.}
\put(77,25){.}
\put(77,25){.}
\put(77,25){.}
\put(76,25){.}
\put(76,25){.}
\put(76,25){.}
\put(75,25){.}
\put(75,25){.}
\put(75,25){.}
\put(74,25){.}
\put(74,25){.}
\put(74,25){.}
\put(73,25){.}
\put(73,25){.}
\put(73,25){.}
\put(72,25){.}
\put(72,25){.}
\put(72,25){.}
\put(71,25){.}
\put(71,25){.}
\put(71,25){.}
\put(70,25){.}
\put(70,25){.}
\put(70,25){.}
\put(69,25){.}
\put(69,25){.}
\put(69,25){.}
\put(68,25){.}
\put(68,25){.}
\put(68,25){.}
\put(67,25){.}
\put(67,25){.}
\put(67,25){.}
\put(66,25){.}
\put(66,25){.}
\put(66,25){.}
\put(65,25){.}
\put(65,25){.}
\put(65,25){.}
\put(64,25){.}
\put(64,25){.}
\put(64,25){.}
\put(63,25){.}
\put(63,25){.}
\put(63,25){.}
\put(62,25){.}
\put(62,25){.}
\put(62,25){.}
\put(61,25){.}
\put(61,25){.}
\put(61,25){.}
\put(60,25){.}
\put(60,25){.}
\put(60,25){.}
\put(59,25){.}
\put(59,25){.}
\put(59,25){.}
\put(58,25){.}
\put(58,25){.}
\put(58,25){.}
\put(57,25){.}
\put(57,25){.}
\put(57,25){.}
\put(56,25){.}
\put(56,25){.}
\put(56,25){.}
\put(55,25){.}
\put(55,25){.}
\put(55,25){.}
\put(54,25){.}
\put(54,25){.}
\put(54,25){.}
\put(53,25){.}
\put(53,25){.}
\put(53,25){.}
\put(52,25){.}
\put(52,25){.}
\put(52,25){.}
\put(51,25){.}
\put(51,25){.}
\put(51,25){.}
\put(50,25){.}
\put(50,25){.}
\put(50,25){.}
\put(49,25){.}
\put(49,25){.}
\put(49,25){.}
\put(48,25){.}
\put(48,25){.}
\put(48,25){.}
\put(47,25){.}
\put(47,25){.}
\put(47,25){.}
\put(46,25){.}
\put(46,25){.}
\put(46,25){.}
\put(45,25){.}
\put(45,25){.}
\put(45,25){.}
\put(325,225){.}
\put(325,225){.}
\put(325,225){.}
\put(319,205){.}
\put(319,220){.}
\put(319,225){.}
\put(318,202){.}
\put(318,219){.}
\put(318,225){.}
\put(317,198){.}
\put(317,218){.}
\put(317,225){.}
\put(316,195){.}
\put(316,217){.}
\put(316,225){.}
\put(315,192){.}
\put(315,216){.}
\put(315,225){.}
\put(309,174){.}
\put(309,211){.}
\put(309,225){.}
\put(308,171){.}
\put(308,210){.}
\put(308,225){.}
\put(307,168){.}
\put(307,209){.}
\put(307,225){.}
\put(306,166){.}
\put(306,208){.}
\put(306,225){.}
\put(305,163){.}
\put(305,207){.}
\put(305,225){.}
\put(299,148){.}
\put(299,202){.}
\put(299,225){.}
\put(298,146){.}
\put(298,201){.}
\put(298,225){.}
\put(297,143){.}
\put(297,200){.}
\put(297,225){.}
\put(296,141){.}
\put(296,199){.}
\put(296,225){.}
\put(295,139){.}
\put(295,199){.}
\put(295,225){.}
\put(289,127){.}
\put(289,193){.}
\put(289,224){.}
\put(288,125){.}
\put(288,192){.}
\put(288,224){.}
\put(287,123){.}
\put(287,191){.}
\put(287,224){.}
\put(286,121){.}
\put(286,191){.}
\put(286,224){.}
\put(285,119){.}
\put(285,190){.}
\put(285,224){.}
\put(279,108){.}
\put(279,184){.}
\put(279,223){.}
\put(278,107){.}
\put(278,183){.}
\put(278,222){.}
\put(277,105){.}
\put(277,183){.}
\put(277,222){.}
\put(276,104){.}
\put(276,182){.}
\put(276,222){.}
\put(275,102){.}
\put(275,181){.}
\put(275,222){.}
\put(269,93){.}
\put(269,175){.}
\put(269,220){.}
\put(268,92){.}
\put(268,174){.}
\put(268,219){.}
\put(267,91){.}
\put(267,173){.}
\put(267,219){.}
\put(266,89){.}
\put(266,172){.}
\put(266,219){.}
\put(265,88){.}
\put(265,171){.}
\put(265,218){.}
\put(259,81){.}
\put(259,165){.}
\put(259,215){.}
\put(258,80){.}
\put(258,164){.}
\put(258,215){.}
\put(257,78){.}
\put(257,163){.}
\put(257,214){.}
\put(256,77){.}
\put(256,162){.}
\put(256,214){.}
\put(255,76){.}
\put(255,161){.}
\put(255,213){.}
\put(249,70){.}
\put(249,154){.}
\put(249,208){.}
\put(248,69){.}
\put(248,153){.}
\put(248,208){.}
\put(247,68){.}
\put(247,152){.}
\put(247,207){.}
\put(246,67){.}
\put(246,151){.}
\put(246,206){.}
\put(245,66){.}
\put(245,149){.}
\put(245,205){.}
\put(239,61){.}
\put(239,142){.}
\put(239,199){.}
\put(238,60){.}
\put(238,141){.}
\put(238,198){.}
\put(237,59){.}
\put(237,140){.}
\put(237,197){.}
\put(236,59){.}
\put(236,139){.}
\put(236,195){.}
\put(235,58){.}
\put(235,137){.}
\put(235,194){.}
\put(229,53){.}
\put(229,130){.}
\put(229,186){.}
\put(228,53){.}
\put(228,128){.}
\put(228,185){.}
\put(227,52){.}
\put(227,127){.}
\put(227,184){.}
\put(226,51){.}
\put(226,126){.}
\put(226,182){.}
\put(225,51){.}
\put(225,124){.}
\put(225,181){.}
\put(219,47){.}
\put(219,116){.}
\put(219,171){.}
\put(218,47){.}
\put(218,115){.}
\put(218,169){.}
\put(217,46){.}
\put(217,114){.}
\put(217,168){.}
\put(216,46){.}
\put(216,112){.}
\put(216,166){.}
\put(215,45){.}
\put(215,111){.}
\put(215,164){.}
\put(209,42){.}
\put(209,103){.}
\put(209,153){.}
\put(208,42){.}
\put(208,101){.}
\put(208,151){.}
\put(207,41){.}
\put(207,100){.}
\put(207,149){.}
\put(206,41){.}
\put(206,98){.}
\put(206,148){.}
\put(205,40){.}
\put(205,97){.}
\put(205,146){.}
\put(199,38){.}
\put(199,89){.}
\put(199,134){.}
\put(198,38){.}
\put(198,88){.}
\put(198,132){.}
\put(197,37){.}
\put(197,86){.}
\put(197,130){.}
\put(196,37){.}
\put(196,85){.}
\put(196,128){.}
\put(195,37){.}
\put(195,84){.}
\put(195,126){.}
\put(189,35){.}
\put(189,76){.}
\put(189,113){.}
\put(188,34){.}
\put(188,75){.}
\put(188,111){.}
\put(187,34){.}
\put(187,73){.}
\put(187,109){.}
\put(186,34){.}
\put(186,72){.}
\put(186,107){.}
\put(185,33){.}
\put(185,71){.}
\put(185,105){.}
\put(179,32){.}
\put(179,64){.}
\put(179,94){.}
\put(178,32){.}
\put(178,63){.}
\put(178,92){.}
\put(177,31){.}
\put(177,62){.}
\put(177,90){.}
\put(176,31){.}
\put(176,60){.}
\put(176,88){.}
\put(175,31){.}
\put(175,59){.}
\put(175,86){.}
\put(169,30){.}
\put(169,53){.}
\put(169,76){.}
\put(168,30){.}
\put(168,52){.}
\put(168,74){.}
\put(167,29){.}
\put(167,51){.}
\put(167,72){.}
\put(166,29){.}
\put(166,50){.}
\put(166,71){.}
\put(165,29){.}
\put(165,50){.}
\put(165,69){.}
\put(159,28){.}
\put(159,44){.}
\put(159,60){.}
\put(158,28){.}
\put(158,44){.}
\put(158,59){.}
\put(157,28){.}
\put(157,43){.}
\put(157,58){.}
\put(156,28){.}
\put(156,42){.}
\put(156,56){.}
\put(155,28){.}
\put(155,41){.}
\put(155,55){.}
\put(149,27){.}
\put(149,38){.}
\put(149,48){.}
\put(148,27){.}
\put(148,37){.}
\put(148,47){.}
\put(147,27){.}
\put(147,36){.}
\put(147,46){.}
\put(146,27){.}
\put(146,36){.}
\put(146,45){.}
\put(145,27){.}
\put(145,35){.}
\put(145,44){.}
\put(139,26){.}
\put(139,33){.}
\put(139,39){.}
\put(138,26){.}
\put(138,32){.}
\put(138,38){.}
\put(137,26){.}
\put(137,32){.}
\put(137,38){.}
\put(136,26){.}
\put(136,31){.}
\put(136,37){.}
\put(135,26){.}
\put(135,31){.}
\put(135,36){.}
\put(129,26){.}
\put(129,29){.}
\put(129,33){.}
\put(128,26){.}
\put(128,29){.}
\put(128,32){.}
\put(127,26){.}
\put(127,29){.}
\put(127,32){.}
\put(126,26){.}
\put(126,29){.}
\put(126,32){.}
\put(125,26){.}
\put(125,28){.}
\put(125,31){.}
\put(119,25){.}
\put(119,27){.}
\put(119,29){.}
\put(118,25){.}
\put(118,27){.}
\put(118,29){.}
\put(117,25){.}
\put(117,27){.}
\put(117,29){.}
\put(116,25){.}
\put(116,27){.}
\put(116,28){.}
\put(115,25){.}
\put(115,27){.}
\put(115,28){.}
\put(109,25){.}
\put(109,26){.}
\put(109,27){.}
\put(108,25){.}
\put(108,26){.}
\put(108,27){.}
\put(107,25){.}
\put(107,26){.}
\put(107,27){.}
\put(106,25){.}
\put(106,26){.}
\put(106,27){.}
\put(105,25){.}
\put(105,26){.}
\put(105,26){.}
\put(99,25){.}
\put(99,25){.}
\put(99,26){.}
\put(98,25){.}
\put(98,25){.}
\put(98,26){.}
\put(97,25){.}
\put(97,25){.}
\put(97,26){.}
\put(96,25){.}
\put(96,25){.}
\put(96,26){.}
\put(95,25){.}
\put(95,25){.}
\put(95,26){.}
\put(89,25){.}
\put(89,25){.}
\put(89,25){.}
\put(88,25){.}
\put(88,25){.}
\put(88,25){.}
\put(87,25){.}
\put(87,25){.}
\put(87,25){.}
\put(86,25){.}
\put(86,25){.}
\put(86,25){.}
\put(85,25){.}
\put(85,25){.}
\put(85,25){.}
\put(79,25){.}
\put(79,25){.}
\put(79,25){.}
\put(78,25){.}
\put(78,25){.}
\put(78,25){.}
\put(77,25){.}
\put(77,25){.}
\put(77,25){.}
\put(76,25){.}
\put(76,25){.}
\put(76,25){.}
\put(75,25){.}
\put(75,25){.}
\put(75,25){.}
\put(69,25){.}
\put(69,25){.}
\put(69,25){.}
\put(68,25){.}
\put(68,25){.}
\put(68,25){.}
\put(67,25){.}
\put(67,25){.}
\put(67,25){.}
\put(66,25){.}
\put(66,25){.}
\put(66,25){.}
\put(65,25){.}
\put(65,25){.}
\put(65,25){.}
\put(59,25){.}
\put(59,25){.}
\put(59,25){.}
\put(58,25){.}
\put(58,25){.}
\put(58,25){.}
\put(57,25){.}
\put(57,25){.}
\put(57,25){.}
\put(56,25){.}
\put(56,25){.}
\put(56,25){.}
\put(55,25){.}
\put(55,25){.}
\put(55,25){.}
\put(49,25){.}
\put(49,25){.}
\put(49,25){.}
\put(48,25){.}
\put(48,25){.}
\put(48,25){.}
\put(47,25){.}
\put(47,25){.}
\put(47,25){.}
\put(46,25){.}
\put(46,25){.}
\put(46,25){.}
\put(45,25){.}
\put(45,25){.}
\put(45,25){.}
\put(25,25){\vector(0,1){210}}
\put(25,25){\vector(1,0){310}}
\put(325,23){\line(0,1){4}}
\put(22,225){\line(1,0){4}}
\put(325,18){$z$}
\put(13,225){$f$}
\put(340,18){$\rho$}
\put(18,240){$distributions$}
\put(0,0){\line(1,0){400}}
\put(0,0){\line(0,1){300}}
\put(0,300){\line(1,0){400}}
\put(400,0){\line(0,1){300}}
\end{picture}

\begin{center}

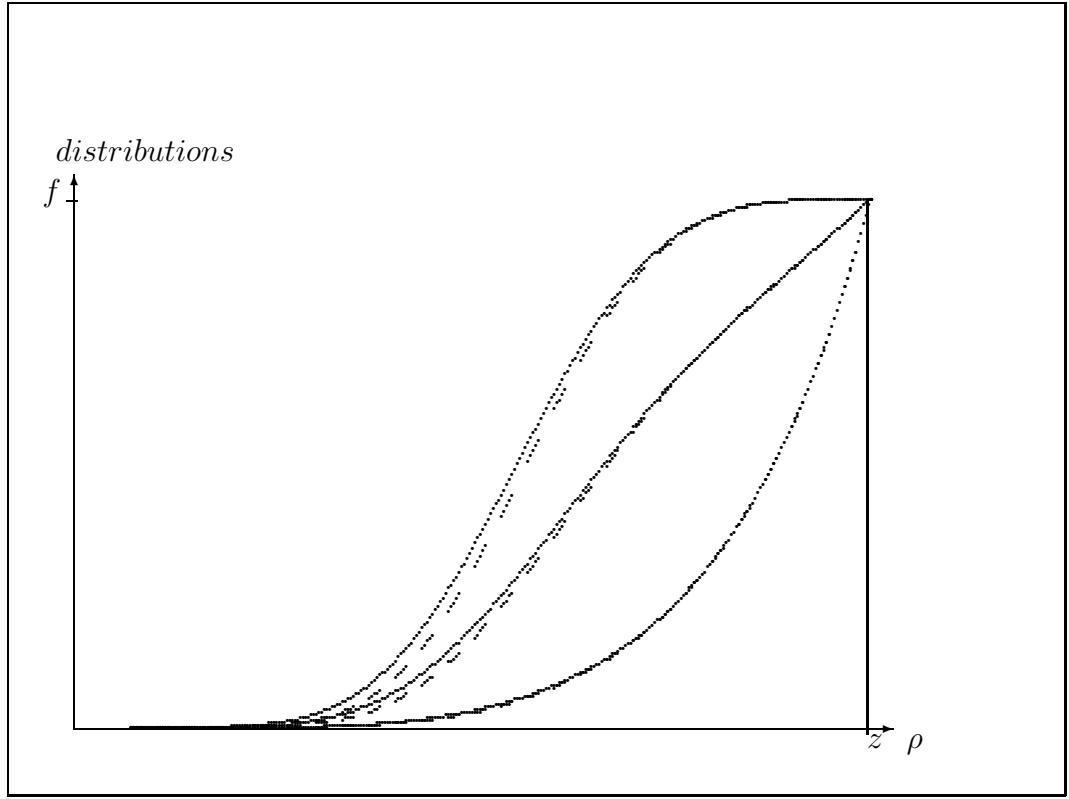
{\it Fig.5}

\end{center}

{\it  Characteristic behavior of size spectrums for nucleation on two
types of heterogeneous centers. One can cancel $f_1$, $\Gamma$ by renormalization.
One can put $f_2 < 1$ due to the choice of centers.
Here $b_1 = 2$, $f_2 =1/2 $, $b_2 = 1/2$. The value $z$ is taken as $
2 \Delta
z $ (see fig.1). The are three curves here. The lower one corresponds
to the spectrum of droplets formed on the first type centers, the intermediate
one
corresponds to the droplets size spectrum for the second type centers,
the upper one corresponds to the spectrum calculated without exhaustion
of heterogeneous centers (the reasons are the same as in fig.3). The solid
lines are the numerical solutions, the dashed lines are the approximate
solutions. One can not see the difference for the lower curve. For the
upper and intermediate curve one can see only very slight difference.
All spectrums are renormalized to have one and the same amplitude (which
is marked by $f$).}

\pagebreak

\begin{picture}(400,300)
\put(152,181){.}
\put(147,175){.}
\put(142,169){.}
\put(137,163){.}
\put(132,158){.}
\put(127,152){.}
\put(122,147){.}
\put(117,141){.}
\put(112,136){.}
\put(107,131){.}
\put(102,125){.}
\put(97,120){.}
\put(92,115){.}
\put(87,109){.}
\put(82,104){.}
\put(77,99){.}
\put(72,94){.}
\put(67,89){.}
\put(62,84){.}
\put(57,78){.}
\put(52,73){.}
\put(158,195){.}
\put(153,188){.}
\put(148,182){.}
\put(143,175){.}
\put(138,169){.}
\put(133,163){.}
\put(128,157){.}
\put(123,151){.}
\put(118,145){.}
\put(113,139){.}
\put(108,133){.}
\put(103,128){.}
\put(98,122){.}
\put(93,117){.}
\put(88,111){.}
\put(83,106){.}
\put(78,100){.}
\put(73,95){.}
\put(68,89){.}
\put(63,84){.}
\put(58,79){.}
\put(165,208){.}
\put(160,200){.}
\put(155,192){.}
\put(150,185){.}
\put(145,178){.}
\put(140,171){.}
\put(135,165){.}
\put(130,158){.}
\put(125,152){.}
\put(120,146){.}
\put(115,140){.}
\put(110,134){.}
\put(105,128){.}
\put(100,122){.}
\put(95,117){.}
\put(90,111){.}
\put(85,105){.}
\put(80,100){.}
\put(75,94){.}
\put(70,89){.}
\put(65,83){.}
\put(172,219){.}
\put(167,210){.}
\put(162,201){.}
\put(157,193){.}
\put(152,186){.}
\put(147,179){.}
\put(142,171){.}
\put(137,165){.}
\put(132,158){.}
\put(127,152){.}
\put(122,145){.}
\put(117,139){.}
\put(112,133){.}
\put(107,127){.}
\put(102,121){.}
\put(97,115){.}
\put(92,109){.}
\put(87,103){.}
\put(82,98){.}
\put(77,92){.}
\put(72,87){.}
\put(178,228){.}
\put(173,218){.}
\put(168,209){.}
\put(163,200){.}
\put(158,192){.}
\put(153,184){.}
\put(148,177){.}
\put(143,169){.}
\put(138,163){.}
\put(133,156){.}
\put(128,149){.}
\put(123,143){.}
\put(118,136){.}
\put(113,130){.}
\put(108,124){.}
\put(103,118){.}
\put(98,112){.}
\put(93,106){.}
\put(88,101){.}
\put(83,95){.}
\put(78,89){.}
\put(185,235){.}
\put(180,224){.}
\put(175,214){.}
\put(170,205){.}
\put(165,196){.}
\put(160,188){.}
\put(155,180){.}
\put(150,173){.}
\put(145,166){.}
\put(140,159){.}
\put(135,152){.}
\put(130,146){.}
\put(125,139){.}
\put(120,133){.}
\put(115,127){.}
\put(110,120){.}
\put(105,114){.}
\put(100,108){.}
\put(95,103){.}
\put(90,97){.}
\put(85,91){.}
\put(192,241){.}
\put(187,229){.}
\put(182,218){.}
\put(177,209){.}
\put(172,200){.}
\put(167,191){.}
\put(162,183){.}
\put(157,175){.}
\put(152,168){.}
\put(147,161){.}
\put(142,154){.}
\put(137,147){.}
\put(132,141){.}
\put(127,134){.}
\put(122,128){.}
\put(117,122){.}
\put(112,116){.}
\put(107,110){.}
\put(102,104){.}
\put(97,98){.}
\put(92,92){.}
\put(198,245){.}
\put(193,232){.}
\put(188,221){.}
\put(183,211){.}
\put(178,202){.}
\put(173,193){.}
\put(168,185){.}
\put(163,177){.}
\put(158,170){.}
\put(153,162){.}
\put(148,155){.}
\put(143,149){.}
\put(138,142){.}
\put(133,136){.}
\put(128,129){.}
\put(123,123){.}
\put(118,117){.}
\put(113,111){.}
\put(108,105){.}
\put(103,99){.}
\put(98,93){.}
\put(205,247){.}
\put(200,234){.}
\put(195,223){.}
\put(190,212){.}
\put(185,203){.}
\put(180,194){.}
\put(175,186){.}
\put(170,178){.}
\put(165,170){.}
\put(160,163){.}
\put(155,156){.}
\put(150,149){.}
\put(145,143){.}
\put(140,136){.}
\put(135,130){.}
\put(130,123){.}
\put(125,117){.}
\put(120,111){.}
\put(115,105){.}
\put(110,99){.}
\put(105,93){.}
\put(212,249){.}
\put(207,235){.}
\put(202,223){.}
\put(197,213){.}
\put(192,203){.}
\put(187,194){.}
\put(182,186){.}
\put(177,178){.}
\put(172,170){.}
\put(167,163){.}
\put(162,156){.}
\put(157,149){.}
\put(152,143){.}
\put(147,136){.}
\put(142,130){.}
\put(137,123){.}
\put(132,117){.}
\put(127,111){.}
\put(122,105){.}
\put(117,99){.}
\put(112,94){.}
\put(218,249){.}
\put(213,235){.}
\put(208,223){.}
\put(203,213){.}
\put(198,203){.}
\put(193,194){.}
\put(188,186){.}
\put(183,178){.}
\put(178,170){.}
\put(173,163){.}
\put(168,156){.}
\put(163,149){.}
\put(158,142){.}
\put(153,136){.}
\put(148,129){.}
\put(143,123){.}
\put(138,117){.}
\put(133,111){.}
\put(128,105){.}
\put(123,99){.}
\put(118,93){.}
\put(225,248){.}
\put(220,234){.}
\put(215,222){.}
\put(210,212){.}
\put(205,202){.}
\put(200,193){.}
\put(195,185){.}
\put(190,177){.}
\put(185,169){.}
\put(180,162){.}
\put(175,155){.}
\put(170,148){.}
\put(165,142){.}
\put(160,135){.}
\put(155,129){.}
\put(150,123){.}
\put(145,117){.}
\put(140,111){.}
\put(135,105){.}
\put(130,99){.}
\put(125,93){.}
\put(232,247){.}
\put(227,233){.}
\put(222,221){.}
\put(217,210){.}
\put(212,201){.}
\put(207,192){.}
\put(202,184){.}
\put(197,176){.}
\put(192,168){.}
\put(187,161){.}
\put(182,154){.}
\put(177,147){.}
\put(172,141){.}
\put(167,134){.}
\put(162,128){.}
\put(157,122){.}
\put(152,116){.}
\put(147,110){.}
\put(142,104){.}
\put(137,98){.}
\put(132,92){.}
\put(238,245){.}
\put(233,231){.}
\put(228,219){.}
\put(223,209){.}
\put(218,199){.}
\put(213,190){.}
\put(208,182){.}
\put(203,174){.}
\put(198,167){.}
\put(193,160){.}
\put(188,153){.}
\put(183,146){.}
\put(178,140){.}
\put(173,133){.}
\put(168,127){.}
\put(163,121){.}
\put(158,115){.}
\put(153,109){.}
\put(148,103){.}
\put(143,97){.}
\put(138,92){.}
\put(245,243){.}
\put(240,229){.}
\put(235,217){.}
\put(230,207){.}
\put(225,197){.}
\put(220,189){.}
\put(215,180){.}
\put(210,173){.}
\put(205,166){.}
\put(200,158){.}
\put(195,152){.}
\put(190,145){.}
\put(185,139){.}
\put(180,132){.}
\put(175,126){.}
\put(170,120){.}
\put(165,114){.}
\put(160,108){.}
\put(155,102){.}
\put(150,97){.}
\put(145,91){.}
\put(52,73){\line(0,-1){28}}
\put(245,243){\line(0,-1){98}}
\put(145,91){\line(0,-1){46}}
\put(52,45){\line(1,0){93}}
\put(245,145){\line(-1,-1){100}}
\put(150,150){\vector(0,1){110}}
\put(150,150){\vector(1,0){110}}
\put(150,150){\vector(-1,-1){110}}
\put(250,148){\line(0,1){4}}
\put(147,250){\line(1,0){4}}
\put(47,50){\line(1,0){4}}
\put(250,140){$6$}
\put(125,250){$0.05$}
\put(25,50){$1$}
\put(265,143){$b$}
\put(160,265){$error$}
\put(20,35){$f$}
\put(0,0){\line(1,0){300}}
\put(0,0){\line(0,1){300}}
\put(0,300){\line(1,0){300}}
\put(300,0){\line(0,1){300}}
\end{picture}

\begin{center}

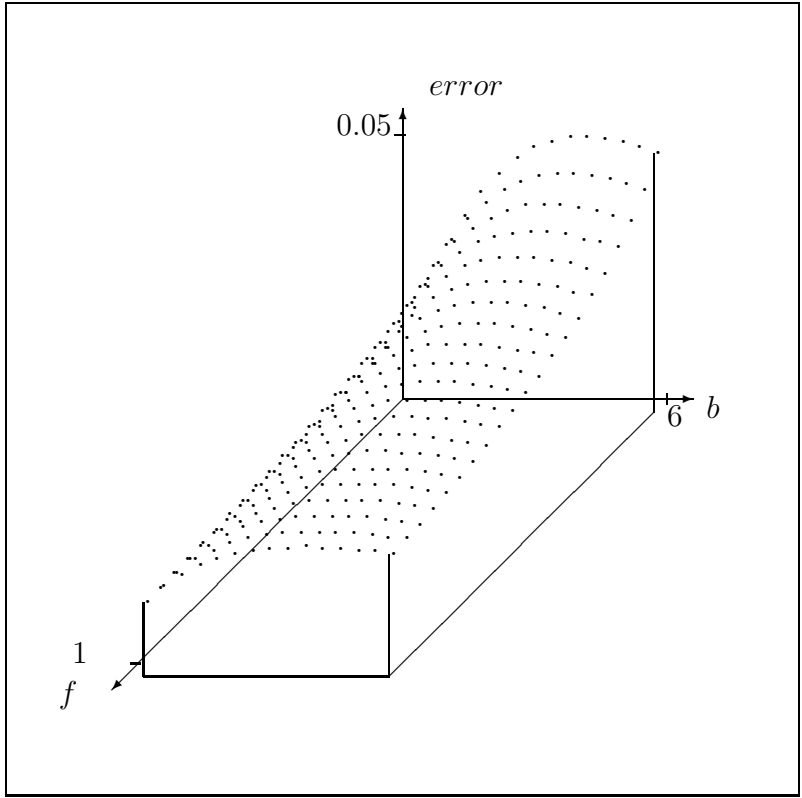
{\it Fig.6}

\end{center}

{\it  Relative error for the nucleation on two types of centers.  For
two types of centers there exists  five parameters (two of them can be
canceled by renormalization). We have already adopt that $\Gamma$ is
one and the same for different types of centers (for the reasons
see \cite{Overlapping},
for numerical results see \cite{Ierarchy}, for analytical estimates see
\cite{iteration}, for recipes of calculation in this situation see \cite{books}).
We cancel here $f_1$, $\Gamma$. We consider $f_2 < 1$ and the first component
is the leading component in the metastable phase consumption. So, the
worst situation for the error in the droplets number formed on the second type
centers (which is drawn here) will be when $b_2=0$ and the is no converging
force of the centers exhaustion. So, there remain two parameters $f_2$
(it is marked by $f$) and $b_1$ (which is marked by $b$). One can see
that the error is small. The calculations for the variations of all parameters
give the same value of error but it is hard to reproduce these numerical
results. }

\end{document}